\documentclass[10pt]{iopart}

\usepackage[flushleft]{threeparttable}
\usepackage{hyperref}
\usepackage{xcolor}
\usepackage{color, soul}
\usepackage{setspace}
\usepackage[normalem]{ulem}
\def\SOUL@hlpreamble{%
    \setul{\dp\strutbox}{\dimexpr\ht\strutbox+\dp\strutbox\relax}%
    \let\SOUL@stcolor\SOUL@hlcolor
    \SOUL@stpreamble
}
\makeatother

\definecolor{yellowlight}{cmyk}{0.0,0.0,1.0,0.0}

\newcommand*\highlightht{2.8ex}
\newcommand*\highlightdp{-.8ex}
\newcommand*\highlightwd{0.2ex}

\newcommand\highlightcommon[1]
  {%
    \bgroup
    \markoverwith
      {\textcolor{#1}{\smash{\rule[\highlightdp]{\highlightwd}{\highlightht}}}}%
    \ULon
  }
\def\cccolorbox#1#2{\ifx#2\relax\let\next\allowbreak\else
       \def\next{\colorbox{#1}{#2}\allowbreak\cccolorbox{#1}}\fi\next}
\def\ccolorbox#1#2{\fboxsep0pt\cccolorbox{#1}#2\relax}

\def\!#1{\ifx#1\ccolorbox\allowbreak\expandafter\ccolorbox\else
         \ifx#1\end\expandafter\expandafter\expandafter\end\else
         #1\allowbreak\expandafter\expandafter\expandafter\!\fi\fi}

\def\cccolorbox#1#2{\ifx#2\relax\let\next\allowbreak\else
   \def\next{\colorbox{#1}{\strut #2}\allowbreak\cccolorbox{#1}}\fi\next}

\usepackage{tikz}

\expandafter\let\csname equation*\endcsname=\relax 
\expandafter\let\csname endequation*\endcsname=\relax 
\usepackage{amsmath,amssymb,amsthm} 
\usepackage{braket} 
  \usepackage{enumerate}
  \usepackage[inline]{enumitem}
  \usepackage{algorithm}
  \usepackage[noend]{algpseudocode}
  \usepackage{makecell}
  \usepackage{array}
  \usepackage{textcomp}
  \usepackage{threeparttable}
  \usepackage{subcaption}
  \usepackage{enumitem}
  \usepackage{array}
  \usepackage{booktabs}
  \usepackage{multirow}
  \usepackage{wrapfig}
  \usepackage{mathtools}
  \usepackage{amssymb}

  \usepackage{dsfont}
  \usepackage{gensymb}
  \usepackage{multirow}
    
    \DeclarePairedDelimiter\floor{\lfloor}{\rfloor}
  \usepackage{graphicx}
  \usepackage{adjustbox}
\usepackage[utf8x]{inputenc}

\makeatletter
\let\MYcaption\@makecaption
\makeatother
\usepackage[font=footnotesize]{subcaption}
\makeatletter
\let\@makecaption\MYcaption
\makeatother

\graphicspath{ {figures/} }

\newcommand{\defcommenter}[2]{%
  \expandafter\newcommand\csname #1\endcsname[1]{%
  {\color{#2}[#1: ##1]}%
  }%
}
\defcommenter{SK}{orange}
\defcommenter{TG}{purple}
\defcommenter{BP}{green}

\providecommand{\keywords}[1]
{
  \small	
  \textbf{\textit{Keywords---}} #1
}

\begin{document}
\title[Training of Quantized Deep Neural Networks Using a MTJ-Based Synapse]{Training of Quantized Deep Neural Networks Using a Magnetic Tunnel Junction-Based Synapse}

\author{Tzofnat Greenberg-Toledo, Ben Perach,
Itay Hubara, Daniel Soudry, Shahar Kvatinsky
\footnote{T. Greenberg-Toledo, Ben Perach, Itay Hubara, Daniel Soudry, and S. Kvatinsky are with the Andrew and Erna Viterbi Faculty of Electrical Engineering, Technion –- Israel Institute of Technology, Haifa, Israel, 3200003. \{stzgrin@tx\},\{daniel.soudry@ee\},\{shahar@ee\}.technion.ac.il}
\ead{stzgrin@campuse.technion.ac.il}
}
\address{Department of Electrical Engineer, Technion - Israel Institute of Technology, Technion City, Haifa  3200003, Isreal.}

\begin{abstract}
Quantized neural networks (QNNs) are being actively researched as a solution for the computational complexity and memory intensity of deep neural networks. This has sparked efforts to develop algorithms that support both inference and training with quantized weight and activation values, without sacrificing accuracy. A recent example is the GXNOR framework for stochastic training of ternary and binary neural networks (TNNs and BNNs, respectively). 
In this paper, we show how magnetic tunnel junction (MTJ) devices can be used to support QNN training. We introduce a novel hardware synapse circuit that uses the MTJ stochastic behaviour to support the quantize update. The proposed circuit enables processing near memory (PNM) of QNN training, which subsequently reduces data movement.
We simulated MTJ-based stochastic training of a TNN over the MNIST, SVHN, and CIFAR$10$ datasets and achieved an accuracy of $98.61\%$, $93.99\%$ and $83.02\%$, respectively (less than $1\%$ degradation compared to the GXNOR algorithm).
We evaluated the synapse array performance potential and showed that the proposed synapse circuit can train TNNs in situ, with $18.3\frac{TOPs}{W}$ for feedforward and $3\frac{TOPs}{W}$ for weight update.
\end{abstract}
\maketitle

\section{Introduction}
Deep neural networks (DNNs) are the state-of-the-art solution for a wide range of applications such as computer vision and natural language processing. The classic DNN approach requires frequent memory accesses and is compute-intensive, requiring numerous multiply and accumulate (MAC) operations.
For example, the ResNet$50$ network requires $3.9$ billion MAC operations, while storing and accessing $25.5$MB of weights~\cite{Proc_Surv}.
As such, DNN performance is limited by computing resources and power budget. Therefore, efforts have been made to design dedicated hardware for DNNs~\cite{DaDianNao,PipeLayer,Soudry2014}. These solutions support training with high resolution, such as $32$-bit floating point.
Still, DNN models are power-hungry and tend not be suitable to run on low-power devices.

Ternary neural networks (TNNs) and binary neural networks (BNNs) are being explored as a way to reduce the computational complexity and memory footprint of DNNs. 
By reducing the weight resolution and activation function precision to quantized binary $\{-1,1\}$ or ternary $\{-1,0,1\}$ values, the MAC operations are replaced by much less demanding logic operations, and the number of required memory accesses is significantly reduced. Such networks are also known as \textit{quantized neural networks} (QNNs)~\cite{Hubara2016}. The potential efficiency of QNNs has motivated research efforts to design novel algorithms that can support BNNs and/or TNNs without sacrificing solution performance (usually measured by prediction accuracy). These efforts include data quantization during training. 
In this work, we focus on the GXNOR training algorithm~\cite{GXNOR}. This algorithm uses a stochastic update function to facilitate the training phase.
Unlike other algorithms~\cite{Courbariaux2016,Courbariau2015,Hubara2016}, GXNOR does not require storing the full value (\textit{e.g.}, in a floating point format) of the weights and activations. Hence, GXNOR enables further reduction of the memory capacity during the training phase.

Emerging memory technologies, such as spin-transfer torque magnetic tunnel junction (STT-MTJ), can be used to design dedicated hardware to support in-situ DNN training, with parallel and energy-efficient operations. The near-memory computation enabled by these technologies also reduces overall data movement. 
The MTJ is a binary device, with two stable resistance states. Switching the MTJ device between resistance states is a stochastic physical process.
While typically, stochastic switching is not a desirable property for memory cells to have, we exploit this feature to support QNN training.

Previous works used the stochastic behavior of the STT-MTJ, or other memristive technologies such as resistive RAM (RRAM), to implement hardware accelerators for BNNs~\cite{XNOR-RRAM,Yu2013,Vincent2015,MTJ_BNN_tr}. In \cite{XNOR-RRAM}, the research focus was on the architecture level of BNN accelerators, without supporting training. Other works implemented hardware for bio-inspired artificial neural networks (ANNs), using the spike-timing-dependent plasticity (STDP) training rule~\cite{Yu2013,Vincent2015}. Although STDP is widely used for bio-inspired ANNs, common DNNs are trained with gradient-based optimization such as stochastic gradient descent (SGD) and adaptive moment estimation (ADAM)~\cite{Ruder}.
A recently proposed MTJ-based binary synapse comprising a single transistor and a single MTJ device ($1$T$1$R)~\cite{MTJ_BNN_tr} supports training QNNs with binary weights and real value activations. \cite{MTJ_BNN_tr} exploited analog computation to support processing near memory (PNM). Their design, however, requires two update operations to execute the SGD updates. Using real-valued activation will require high-resolution data converters, thereby increasing the area and power consumption of the proposed solution.

In this paper, we explore the stochastic behavior of the MTJ and leverage it to support fully quantized training (GXNOR). Our solution reduces the overall weight and read operations and the cost of the update phase.
We propose a four-transistor, two-MTJ ($4$T$2$R) circuit for a ternary stochastic synapse and a two-transistor, single-MTJ ($2$T$1$R) circuit for a binary stochastic synapse, where the intrinsic stochastic switching behavior of the MTJ is used to perform a stochastic update function.
Such a design enables highly parallel, energy-efficient, and accurate in-situ computation.
Our designed synapse can support various DNN optimization algorithms, such as SGD and ADAM, which are used regularly in practical applications. 

We evaluated TNN and BNN training using the proposed MTJ-based synapse with PyTorch over the MNIST~\cite{MNIST}, SVHN~\cite{SVHN},and CIFAR$10$~\cite{CIFAR10} datasets, where the circuit parameters were extracted from SPICE simulations using a GlobalFoundries $28$nm FD-SOI process. 
Our results show that using the MTJ-based synapse for training yielded similar results as the ideal GXNOR algorithm, with a small accuracy loss of $0.7\%$ for the TNN and $2.4\%$ for the BNN.

This paper makes the following contributions. It
\begin{itemize}
  \item Exploits the MTJ stochastic properties to support QNN stochastic training.
  \item Demonstrates MTJ applicability within the GXNOR framework. We show that PNM of stochastic QNN training is enabled using the MTJ-based synapse, with only a small accuracy reduction.
  \item Offers MTJ-based ternary and binary synapse circuits. These circuits:
  \begin{itemize}
      \item Exploit the stochastic switching of the MTJ device to support a stochastic weight update algorithm,
      \item Support in-situ weight update of standard optimization algorithms such as SGD and ADAM, without reading the weight data out of the synapse array,
      \item Support near-memory processing of the feedforward and backpropagation computations, enabling high parallelism.
  \end{itemize}
\end{itemize}

The rest of the paper is organized as follows. In Sections~\ref{sec:Pre}, background on DNN and QNN training and MTJ is given. Section~\ref{sec:motivation} addresses the motivation of MTJ-based training.
Section~\ref{sec:MTJ_TNN_syn} describes the proposed MTJ-based ternary synapse.
In Section~\ref{sec:eval}, the ability of the proposed circuits to support TNN training is evaluated as well as  their energy efficiency. In Section~\ref{sec:comp_prev_works} a comparison to previous works is given. A conclusion is provided in Section~\ref{sec:conclusion}.
In the supplementary we explain how to modify the proposed circuits to support BNNs.

\section{Preliminaries} \label{sec:Pre}
\subsection{Deep Neural Networks} \label{ssec:DNN}
DNNs are machine learning models, that use connected layers, composed of neurons, to learn a desired functionality $\mathcal{F}$. The different neuron layers are connected through weighted connections called synapses.
For simplicity, this section focuses on a fully connected (FC) layer; however, a similar computation is done for layers with other weight connections, such as convolution layers (CONV)~\cite{DaDianNao,Courbariaux2016,Courbariau2015,ISAAC}. 
In FC, the output is given by a matrix-vector multiplication
\begin{equation} \label{eq:SGD2}
  \vec{o} = W\vec{x},
\end{equation}
where the elements of matrix $W$ are the synapse weights, $\vec{x}$ is the input neuron vector and $\vec{o}$ is the output. Hence, each element in the output vector $o_m$ is the weighted sum of the input, 
\begin{equation} \label{eq:SGD}
  o_m = \sum_{n=1}^{N}w_{mn}x_n,
\end{equation}
where $N$, $o_m$, $w_{mn}$, and $x_n$ are, respectively, the number of input neurons, the output $m$, the synapse weights between neuron $m$ and neuron $n$, and the value of input neuron $n$.
The following neuron layer is computed by passing $\vec{o}$ through a non-linear function, called an activation function $\sigma(\cdot)$ and is, therefore, given by
\begin{equation} \label{eq:SGD3}
  \vec{x}^{(l+1)} =\sigma(\vec{o}^{(l)}) = \sigma(W^{(l)}\vec{x}^{(l)}),
\end{equation}
where $l$ is the layer index.

\subsection{Training a DNN} \label{ssec:SGD}
\label{sec:NN}
In supervised learning, the network is trained to find the set of parameters, \textit{i.e.}, synapse weights, which approximates the desired functionality. The network is trained using a dataset $\mathcal{D}=\{\vec{x_i}^{(0)},\vec{d_i}\}_{i=1}^n$, where $\vec{x_i}^{(0)}$ is the input vector of the network and $\vec{d_i}$ is the desired output. 
During training, the network parameters $w$ are calibrated to find the desired relation $\vec{d}=\mathcal{F}(\vec{x}^{(0)},W)$. To this end, a measure of quality is defined: the cost function $C(\vec{d},\vec{O})$, where $\vec{O}=\mathcal{F}(\vec{x}^{(0)},w)$ is the output of the network.
The goal of the learning algorithm is to find $w$ that minimizes the value of $C(\vec{d},\vec{O})$ with respect to the dataset. Hence, optimization algorithms are used to find the minimum of $C$. Different optimization algorithms, such as SGD and ADAM, are used during DNN training~\cite{Ruder}. 
During training, first the output and the cost function are computed in a stage called \textit{feedforward}. After the cost function is known, the error of each layer $y^l$ is computed in a stage called \textit{backpropagation}. The error is computed using the chain rule and is given by
\begin{equation} \label{eq:SGD_FC2}
  \vec{y}^{(l)} \overset{\Delta}{=} ((W^{(l+1)})^T\vec{y}^{(l+1)})\cdot{\sigma' (\vec{o}^{(l)})},
\end{equation}
where $W^T$ is the transpose of the weight matrix $W$, and $\sigma'$ is the derivative of $\sigma$ with respect to $\vec{o}$.
Taking the computed error of the layer, the weight gradients are computed and used to update the weights. Usually, the weight update rule is given by
\begin{equation} \label{eq:SGD_FC}
  W^{(l)} = W^{(l)} + f_{opt}(a^l,y^l,W^l), 
\end{equation}
where $f_{opt}$ is defined by the optimization algorithm that is used.

General DNNs do not limit the value of the weights, which can be any real value. Typically, the unconstrained parameters are represented by precision higher than $1$ or $2$-to $32$-bit floating point. The following section describes a framework to train QNNs.

\subsection{Training Quantized Neural Networks with a Stochastic Update Rule}\label{sec:GXNOR}
In recent years, efforts have been made to make DNN models more hardware-compatible. Quantization methods have been explored, where the DNN weights and activation functions are constrained to being discrete values such as binary $\{-1,1\}$ or ternary $\{-1,0,1\}$ values.
For BNNs and TNNs, MAC operations are replaced with the simpler XNOR or Gated-XNOR (GXNOR) logic operations, respectively. The memory footprint of the quantized network is dramatically reduced (for example, for ResNet$50$, with ternary weights and activations, the memory capacity is cut in half during training and by $16$ during inference).

This section describes the GXNOR framework~\cite{GXNOR} that constrains the weights and activations to the quantized space while training the QNN.
We focus on the differences between the GXNOR training algorithm and regular DNN training.

\subsubsection{Quantized Weights and Activations}
The quantized space $Z_N$ is defined by
\begin{equation}
    Z_N = \{\frac{n}{2^{N-1}}-1|n=0,1...,2^N\} \in [-1,1],
\end{equation}
where $N$ is a non-negative integer that defines the space values. For example, the binary space is given for $N=0$ and the ternary space for $N=1$. The quantized space resolution, \textit{i.e.}, the distance between two adjacent states, is given by
\begin{equation}
    \Delta z_N = \frac{1}{2^{N-1}}.
\end{equation}

\subsubsection{Feedforward and Backpropagation} 
In QNNs, the quantized activation function is a step function, where the number of steps is defined by the space. To support backpropagation through the quantized activations ($\varphi_r$), the derivative of the activation function is approximated. In this work, the ideal derivative is approximated by a sum of window functions. The window function is given by
\begin{equation}
    \frac{\partial\varphi_r(x)}{\partial x} = 
    \begin{cases}
        \frac{1}{2a},\ if\ r-a\leq x \leq r+a,\\
        0,\ others
    \end{cases}
\end{equation}
where $r$ and $a$ are positive hyperparameters, defining the sparsity of the neurons (\textit{i.e.}, the quantization range) and the window function width, respectively.
Using the approximated derivative, the backpropagation of the GXNOR training algorithm is computed with no further changes compared to regular DNNs.

\subsubsection{Weight Update}
\label{ssec:Gxnor_Wup}
To support training with weights constrained to the discrete weight space (DWS), the GXNOR algorithm uses a stochastic gradient-based method to update the weights. First, the update value is computed by an optimization algorithm (\textit{e.g.}, SGD, ADAM, RMSprop). Then, a boundary function is defined to guarantee that the updated value will not exceed the $[-1,1]$ range. The boundary function is 
\begin{equation}\label{eq:bound}
\begin{aligned}
&\varrho(W^l_{ij}(k),\Delta W^l_{ij}(k))=
&\begin{cases}
min(1- W^l_{ij}(k),\Delta W^l_{ij}(k)),\ & if \ \Delta W^l_{ij}(k)>0,\\
max(-1- W^l_{ij}(k),\Delta W^l_{ij}(k)),\  & else
\end{cases}
\end{aligned}
\end{equation}
where $W^l_{ij}\in Z_N$ is the synaptic weight between neurons $j$ and $i$ of the following layer ($l+1$), $\Delta W^l_{ij}\in \mathbb{R}$ is the gradient-based update value, and $k$ is the update iteration. 
Then, the update step is given by
\begin{equation}\label{eq:sto_update}
 W_{ij}^l(k+1)=W_{ij}^l(k)+\Delta w^l_{ij}(k),
\end{equation}
where $\Delta w^l_{ij}(k)=\mathcal{P}(\varrho)\in\mathbb{Z}$ is the discrete update value, obtained by projecting $\varrho( \Delta W^l_{ij}(k))$ to a quantized weight space. $\mathcal{P}(\varrho)$ is a probabilistic projection function defined by
\begin{equation}\label{eq:prob_proj}
\mathcal{P}(\varrho) = 
\begin{cases}
\kappa_{ij}\Delta z_N+sign(\varrho)\Delta z_N, & w.p. \ \eta\big(\nu_{ij}\big) \\
\kappa_{ij}\Delta z_N, & w.p. \ 1-\eta\big(\nu_{ij}\big),
\end{cases}
\end{equation}
where $\kappa_{ij}$ and $\nu_{ij}$ are, respectively, the quotient and remainder values of $\varrho$ divided by $\Delta z_N$, and
\begin{equation} \label{eq:nu}
\eta\big(\nu\big)=tanh\Bigg(m\cdot\frac{|\nu|}{\Delta z_N}\Bigg)\in[0,1],
\end{equation}
where $m$ is a positive hyperparameter. Hence, 
\begin{equation} \label{eq:w_up}
\Delta w^l_{ij}=\kappa_{ij}\Delta z_N + sign(\nu_{ij})Bern(\eta(\nu_{ij}))\Delta z_N,
\end{equation} 
where $Bern(\eta(\nu_{ij}))$ is a Bernoulli random variable with parameter $\eta(\nu_{ij})$. 

\begin{figure}[!t]
    \centering
    \includegraphics[trim={0cm 0.5cm 0cm 0cm},width=0.7\columnwidth]{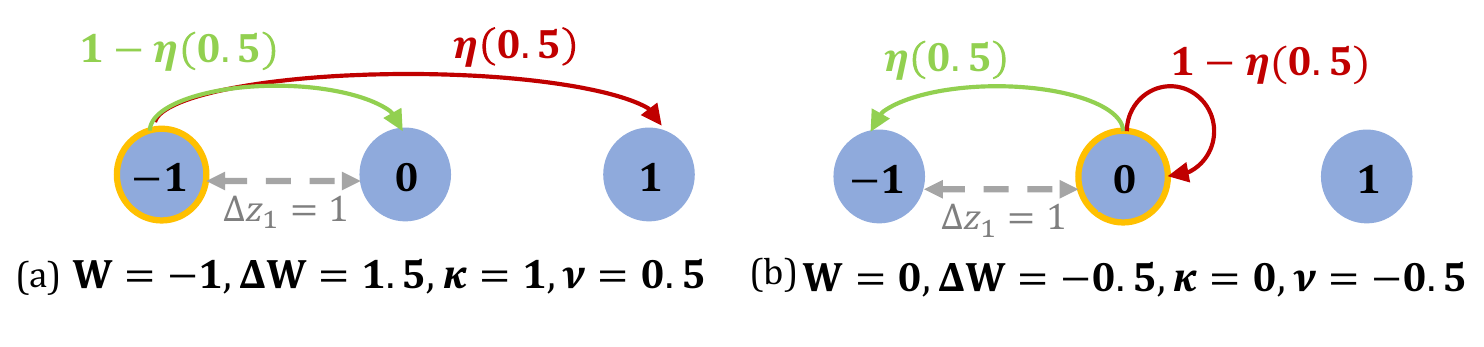}
    \caption{TNN examples (TWS): Ternary weight update with $\Delta z_1=1$. (a) Given $W=-1$, $\Delta W=1.5$, $\kappa=1$, and $\nu=0.5$, the discrete update value is $\Delta w^l=1 + Bern(\eta(0.5))$. (b) For $W=0$, $\Delta W=-0.5$, $\kappa=0$, and $\nu=-0.5$, the discrete update value is $\Delta w^l=-Bern(\eta(-0.5))$.}\label{fig:examples_comb}
\end{figure}

In this paper, which focuses on TNN, the ternary weight space (TWS) is given by $N=1$ and $\Delta z_1 = 1$. Figure~\ref{fig:examples_comb} illustrates examples of TNN weight updates for $W=-1$ and $W=0$.
Further discussion of the BNN implementation is found in the supplementary material.

Dedicated hardware for TNN and BNN can fully exploit the potential of these networks. In this paper, we propose to use emerging memory technology, \textit{i.e.}, STT-MTJ, to support PNM of TNNs and BNNs.

\subsection{Magnetic Tunnel Junction}\label{sec:MTJ}
An MTJ device comprises two ferromagnetic layers, a fixed magnetization layer and a  free magnetization layer, separated by an insulator layer,  as illustrated in Figure~\ref{fig:tsyn_simplified}. The resistance of the device is defined by the relative magnetization of the free layer as compared to the fixed layer. A parallel magnetization state (P) leads to low resistance ($R_{on}$) and an anti-parallel state (AP) leads to high resistance ($R_{off}$). The device resistance can be switched by the current flow through the device. When the current flows from the free layer to the fixed layer, the resistance may switch to $R_{on}$. Likewise, when the current flows from the fixed layer to the free layer, the resistance may switch to $R_{off}$.
The switching probability of the MTJ device depends on the current's magnitude, when three work regimes are defined as: 1) low current, low switching probability, 2) intermediate current, and 3) high current, high switching probability~\cite{Vincent2015}. 
As we are interested in fast switching time, this work focuses on the high current regime. Therefore, current $I$ is substantially higher than critical current $I_{c_0}$, and is given by
\begin{equation} 
 I_{c_0} = \frac{2|e|}{\hbar} \frac{\alpha V(1\pm P)}{P} \mu_0 M_s \frac{M_{eff}}{2},
\end{equation}
where $\alpha$, $M_s$, $V$, $P$, $M_{eff}$ are, respectively, the Gilbert damping, the saturation magnetization, the free layer volume, the spin polarization of the current, and the effective magnetization~\cite{Analitic_MTJ}.
The switching time is, therefore,
\begin{equation}\label{eq:T_swHC}
 \tau = \frac{2}{\alpha\gamma\mu_0M_s}\frac{I_{c_0}}{I-I_{c_0}}log\Big(\frac{\pi}{2|\theta|} \Big),
\end{equation}
where $\gamma$ is the gyromagnetic ratio, and $\theta$ is the initial magnetization angle~\cite{Analitic_MTJ}, given by a normal distribution $\theta \sim \mathcal{N}(0,\theta_0)$, $\theta_0=\sqrt{k_BT/(\mu_0H_kM_sV)}$, where $H_k$, $K_b$ and $T$ are the shape anisotropy field, the Boltzmann constant and temperature, respectively.
In this work, we use current pulses with varying time intervals to control the MTJ switching probability and to support the stochastic weight update given by (\ref{eq:w_up}).

\section{Stochastic In-Situ Training}~\label{sec:motivation}
The training scheme suggested in~\cite{GXNOR} reduces the memory footprint of the training phase. Still, every update iteration (Eq.~(\ref{eq:sto_update})) requires reading the weights, computing the stochastic update step, and writing the new weight value. Computing the stochastic update will require the use of a random number generator (RNG). Adding RNG will increase the complexity of the design, in terms of having to transfer the random numbers to all the weights, area overhead of the RNG circuit and the resulting power consumption. For example, assume $128$ weights are read in $100$ns~\cite{ISAAC}, to generate $128$ $8$-bit random numbers at this rate, the RNG design requires $64$ PRNG circuits~\cite{PRNG}.
{\color{black}We suggest replacing the RNG functionality with the stochastic write operation of the MTJ device}. Our approach replaces the read-PRNG-write loop with a single stochastic-write operation.  {\color{black}Moreover, working with the MTJ in a stochastic write regime allows us to work with shorter write intervals.}
Other emerging memory technologies also have stochastic write models that might be a good fit to the expression in~(\ref{eq:nu}). Nevertheless, training requires numerous write operations; {\color{black}for example, one network training with $1000$ training epochs requires $5\cdot 10^7$, and $10^8$ writes per device for CIFAR-$10$ and ImageNet datasets, respectively.} Thus, a STT-MTJ device, which has the high reported endurance, is a leading candidate for stochastic in-situ training~\cite{Endurance}. 

\section{MTJ-Based Ternary Synapses}\label{sec:MTJ_TNN_syn}
We now describe the proposed ternary synapse circuit that supports stochastic GXNOR training.
In the supplementary, we explain how the proposed synapse can support binary weights as well.

\subsection{Training TNN Using an MTJ-Based Synapse}~\label{ssec:stoch_wup}
First, we describe how we leverage the stochastic switching behavior of the MTJ device to perform the stochastic update function.
Two MTJ devices are needed to represent ternary weight, where the weight is defined and stored as the combination of the resistances of the two MTJs. Table~\ref{tab:syn_state} lists the different values of the synapse weight as a function of the MTJ's resistance. To support the stochastic weight update, both MTJ devices might be switched during an update. 
To switch the state of the MTJ device, a voltage pulse $V_{up}$ is applied across the device, for time interval $\Delta t\in [0,T_{up}]$. For a fast update operation, the update is performed in the high current domain guaranteed by $V_{up}$. The resultant current direction and the pulse time interval determine the switching probability. Using (\ref{eq:T_swHC}) and the voltage pulse, the switching probability of the MTJ is
\begin{equation} \label{eq:p_sw}
 P_{sw}(\Delta t) = 1-erf\Bigg(\frac{\pi}{2\sqrt{2}\theta_0\exp{\Big(\frac{\Delta tV_{up}}{CR}\Big)}} \Bigg),
\end{equation}
where $C=\frac{2I_{c_0}}{\alpha\gamma\mu_0M_s}$, and $R$ is the device resistance. As indicated in Eq.~(\ref{eq:p_sw}), $P_{sw}$ is a function of the voltage pulse amplitude and time interval. Therefore, $T_{up}$ is set to guarantee that if $\Delta t = T_{up}$, then $P_{sw}\approx 1$. Moreover, $P_{sw}$ is a function of the current direction flows through the MTJ {\color{black}and the state of the MTJ device. }
\begin{table}[t!]
    \centering
    \caption{Ternary Synapse States and Output Current}
    \renewcommand{\arraystretch}{1}
    \begin{tabular}{|c|c|c|c|}
    \hline
        \textbf{Weight} & \boldmath{$R_1$} & \boldmath{$R_2$} & \boldmath{$I_{out}$} \\
                    \hline\hline
        $1$   & $R_{on}$  & $R_{off}$  & $\frac{R_{off}-R_{on}}{R_{off}R_{on}}u$ \\
                    \hline
        $0_s$ & $R_{off}$  & $R_{off}$ & $0$ \\
                    \hline
        $0_w$ & $R_{on}$  & $R_{on}$   & $0$ \\
                    \hline        
        $-1$  & $R_{off}$  & $R_{on}$  & $-\frac{R_{off}-R_{on}}{R_{off}R_{on}}u$ \\
                    \hline    
    \end{tabular}
    \label{tab:syn_state}
\end{table}

\begin{figure}[!t]
    \centering
    \includegraphics[width=0.4\linewidth]{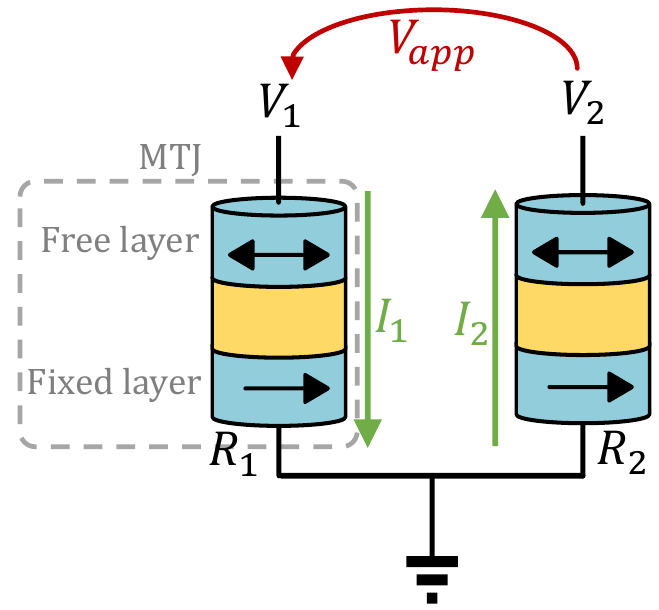}
    \caption{Simplified model of the ternary synapse. The synapse weight is define by the resistance combination of $R_1$ and $R_2$. As listed in Table~\ref{tab:syn_state} each weight has four possibles values $w\in \{-1, 0_s, 0_w, 1\}$.}\label{fig:tsyn_simplified}
\end{figure}

To better understand the update operation of a single synapse, in this section we consider the simplified synapse illustrated in Figure~\ref{fig:tsyn_simplified}.
Each MTJ update is independent; this is guaranteed by applying different voltage pulses $V_{1},V_{2}$ on each synapse and connecting the node between the MTJs to the ground. In this manner, each MTJ is updated according to (\ref{eq:p_sw}). 
To support the GXNOR update, we need to control the switching probability of each MTJ device according to the update value $(\Delta W)$ and the synapse weight. To this end, we
\begin{enumerate*}
    \item define $V_{app} = V_1-V_2$,
    \item enforce opposite polarities of $V_{1}$ and $V_{2}$ (\textit{i.e.}, $sign(V_1)\ne sign(V_2)$), and
    \item set $sign(V_{app})=sign(\Delta W)$.
\end{enumerate*}

Following this work scheme, if $\Delta W>0$, the current directions guarantee that only a synapse with weight $W=-1$ or $W=0_s$ can switch. Similarly, if $\Delta W<0$, the current directions guarantee that only a synapse with weight $W=1$ or $W=0_w$ can switch. 

Next, we need to ensure that the switching probability will follow Eq.~(\ref{eq:w_up}) and will be a function of the update value $\Delta w=\kappa + sign(\nu)Bern(\eta(\nu))$.
As indicated by Eq.~(\ref{eq:p_sw}), the pulse duration $\Delta t$ sets the switching probability. To support (\ref{eq:w_up}), we set the pulse duration of $V_{1}$ and $V_{2}$ to be a function of $\kappa$ or $\nu$, where $\kappa=\{0,1,2\}$ and $\nu \in [0,1]$.
 If $\Delta W >0$, the pulse duration of $V_1$ is set by $\kappa$. Hence, $\Delta t_{V_1}=f(\kappa)=\mathds{1}_{\kappa\ne0}T_{up}$. The pulse duration of $V_2$ is set by $\nu$; so, $\Delta t_{V_2} = f(\nu)=\nu T_{up}$.
Similarly, if $\Delta W<0$, then $\Delta t_{V_1}=f(\nu)$ and $\Delta t_{V_2}=\mathds{1}_{\kappa\ne0}T_{up}$.

Following this methodology, at each weight update, one MTJ is updated as a function of $\kappa$, while the other is updated as a function of $\nu$, depending on the sign of $\Delta W$. 
Thus, if $\kappa\ne0$, the MTJ switching probability is approximately $1$ and the switching probability is given by the indicator variable $P_{sw,\kappa}=\mathds{1}_{\kappa\ne0}$.
Since $\nu$ is a fraction, the switching probability of the other MTJ with respect to $\nu$ is a Bernoulli variable with probability $P_{sw,\nu}=P(\nu T_{up})$. 
Therefore, the MTJ-based synapse update is given by 
\begin{equation}
\begin{aligned}
    \Delta w=&sign(\Delta W)(P_{sw,\kappa}+P_{sw,\nu})= 
    &sign(\Delta W)(\mathds{1}_{\kappa\ne0}+Bern(P(\nu T_{up}))); 
\end{aligned}    
\end{equation}
see examples in Section~\ref{ssec:tnn_exp}. 

The MTJ-based synapse update differs from the ideal GXNOR update in that it supports two zero states, and uses similar, but not identical, switching probabilities ($P_{sw}\approx \eta$).

\subsection{Proposed Synapse Circuit and Synapse Array}
\subsubsection*{Synapse Circuit}\label{sssec:syncir}
A schematic of the proposed ternary synapse is shown in Figure~\ref{fig:synapse}. The ternary synapse is composed of two MTJ devices connected via their fixed layer port. The free layer port of each MTJ is connected to two access transistors. This synapse is inspired by previous work~\cite{Soudry2014,tz_mom}, but we replace the RRAM by the MTJ device, and two synapse structures are added together to support the ternary weight. In contrast to~\cite{Soudry2014,tz_mom} which supports full-precision analog weight values, the MTJ-based synapse supports quantized weights and stochastic weight updates. Sections~~\ref{ssec:MTJ-TNN_Wup} and \ref{ssec:TNN_FFnBP} describe how our design is optimized to support quantized weights.

\subsubsection*{Synapse Array}
\begin{figure*}[!t]
        \begin{subfigure}[b]{0.15\textwidth}
                \includegraphics[trim={9cm 0cm 6cm 3cm},width=1\linewidth]{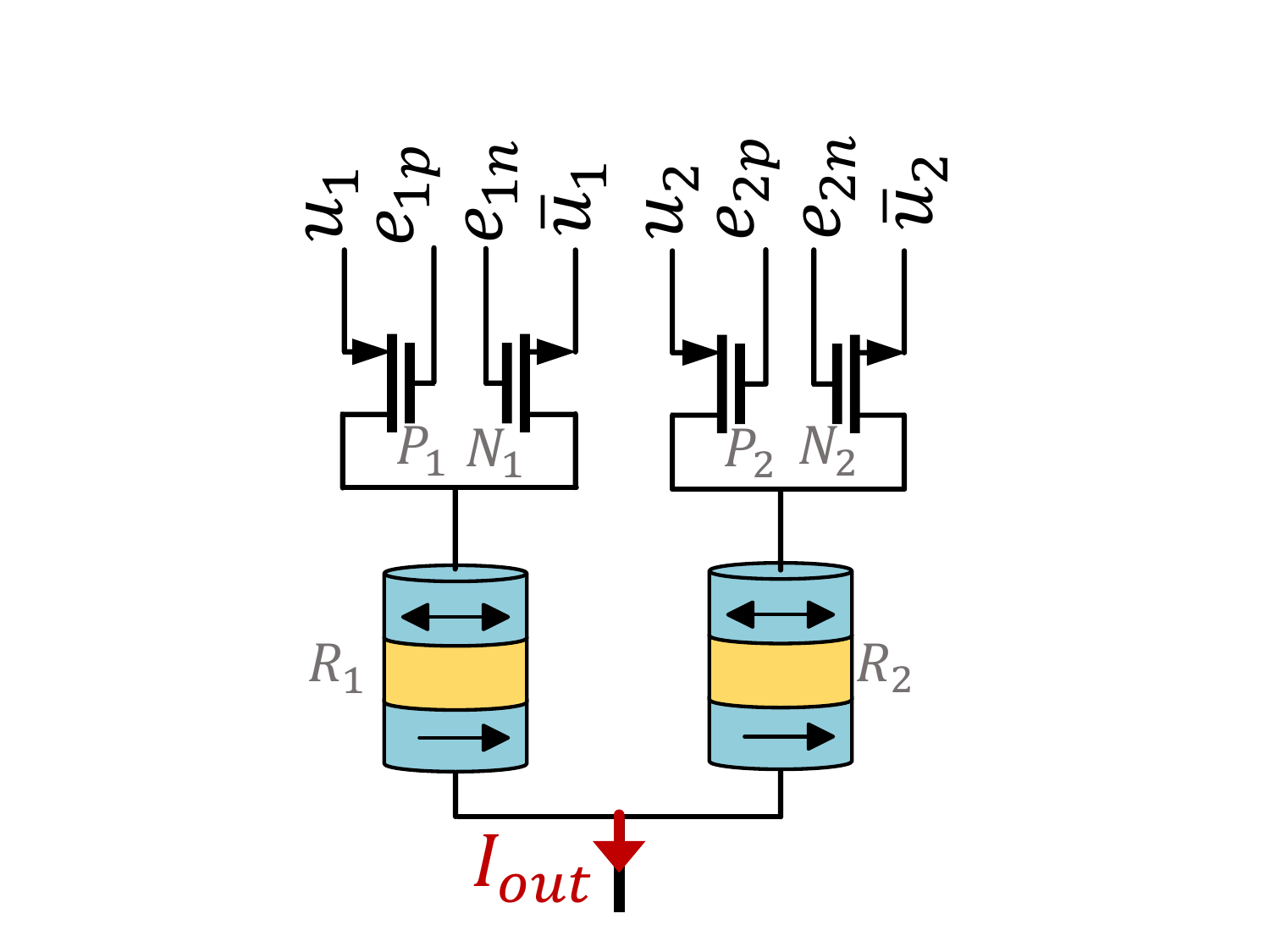}
                \caption{\small Synapse cell}
                \label{fig:synapse}
        \end{subfigure}\hspace{1ex}%
        \begin{subfigure}[b]{0.5\textwidth}
                \includegraphics[trim={0cm 2cm 0cm 0cm},width=0.9\linewidth]{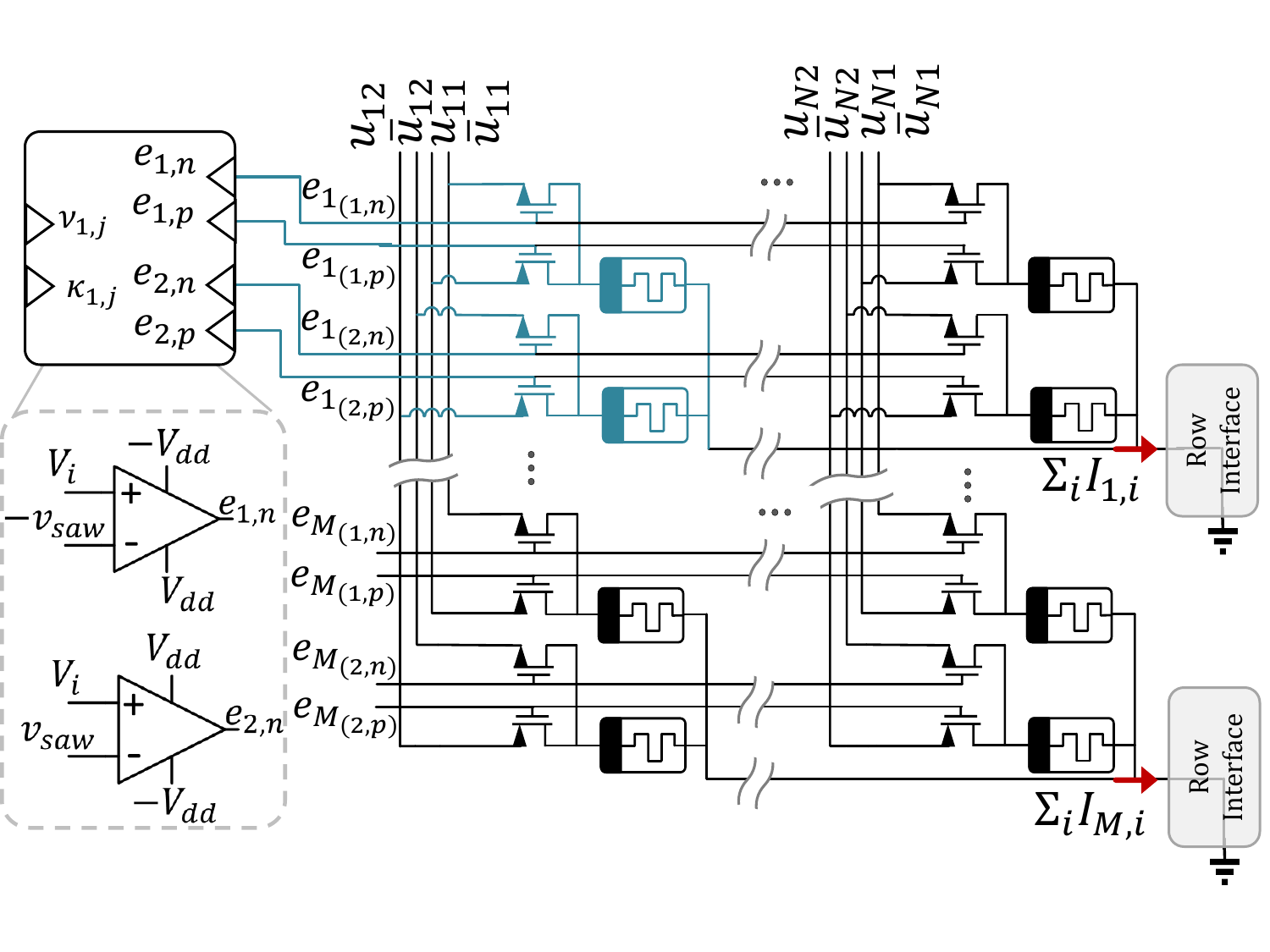}
                \caption{\small Synapse array}
                \label{fig:array}
        \end{subfigure}\hspace{0.5ex}%
        \begin{subfigure}[b]{0.3\textwidth}
                \includegraphics[trim={2cm 0.5cm 5cm 0cm},width=0.95\linewidth]{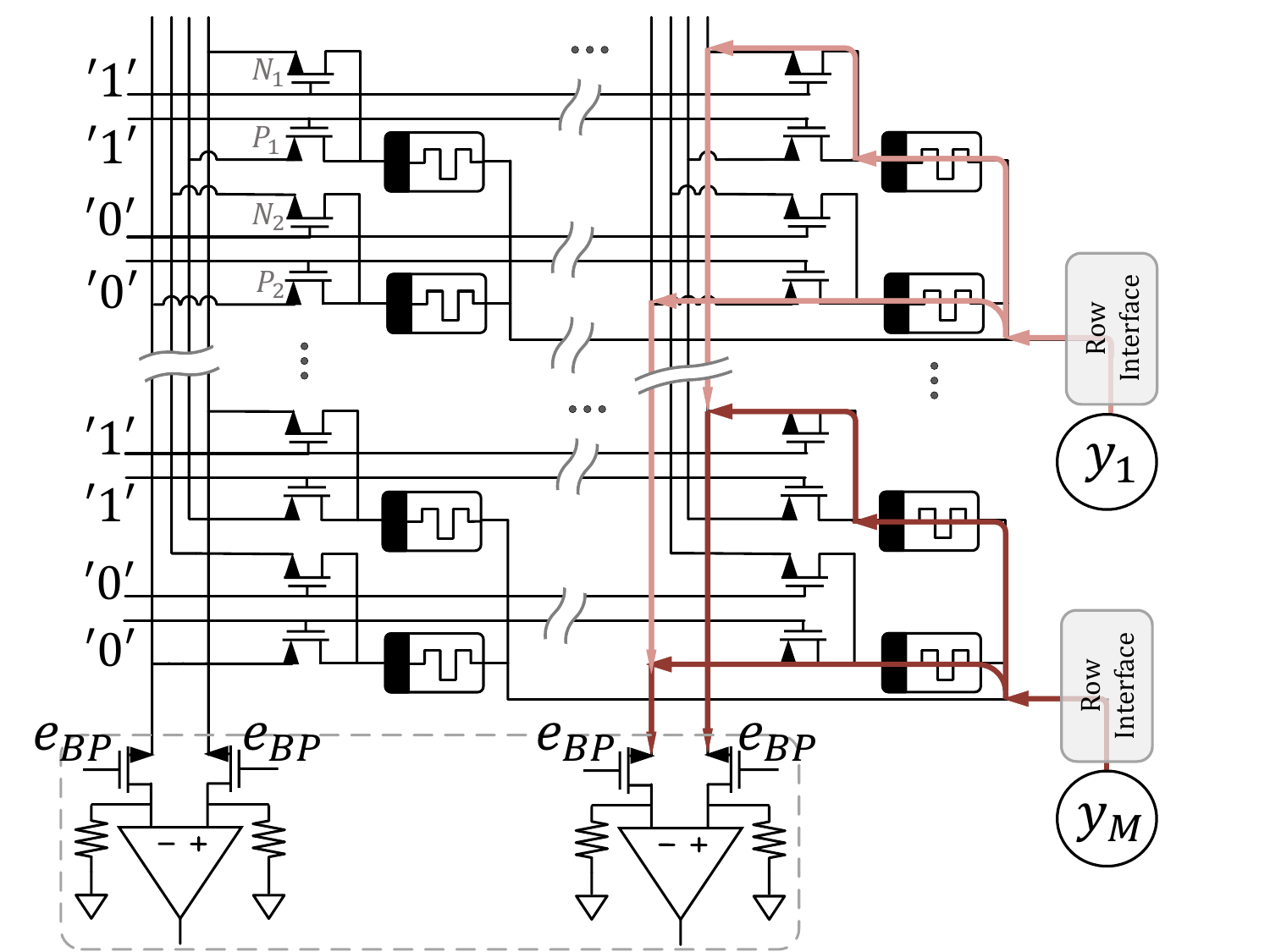}
                \caption{\small Backpropagation}
                \label{fig:inv_rd}
        \end{subfigure}%
        \caption{(a) Schematic of the 4T2R MTJ-based synapse. (b) The synapse array. The MTJ symbol is replaced by the general memristor symbol. An example of a single synapse cell is marked in blue. The control signals are generated using a voltage comparator. (c) The data flow of the backpropagation operation. In the figure $'1'=V_{dd}$, $ '0'=-V_{dd}$.
       }\label{fig:Circuit}
\end{figure*}

The synapse circuit shown in Figure~\ref{fig:synapse} is the basic cell of an array structure, as shown in Figure~\ref{fig:array}. The synapses are arranged in an $M\times N$ array, where each synapse is indexed by $(m,n)$. Each synapse in column $n\in[1,N]$ is connected to four inputs $\{u_{n1},\Bar{u}_{n1},u_{n2},\Bar{u}_{n2}\}$, where all the input voltages are shared among all synapses in the same column. Likewise, each synapse in row $m\in[1,M]$ is connected to control signals $\{e_{m(1,n)},e_{m(1,p)},e_{m(2,n)},e_{m(2,p)}\}$. The control signals are shared among all synapses in the same row. The synapse located in $(m,n)$ produces an output current $I_{mn}$, which contributes to the current through output row $m$. The operations on the synapse array are performed in the analog domain and accumulate according to Kirchoff's current law (KCL), where the GXNOR output is represented by the current.

\subsection{Stochastic Weight Update}
\label{ssec:MTJ-TNN_Wup}
We now explain how the synapse circuit is designed and how the input and control signals are set to support the GXNOR stochastic update scheme.
Unlike weight updates in standard DNN, the proposed synapse supports the quantized update scheme suggested in~\cite{GXNOR}.

\subsubsection{Weight Update Step}
Similar to~\cite{Soudry2014,Eyal_Sergey}, four transistors are added to support parallel synapse updates. 
The control signals of these transistors dictate the weight update functionality by controlling the current direction and the voltage pulse time interval in the synapse array. The update step can be performed in parallel for all the synapses in the same array, depending on the optimization algorithm used.
Since the GXNOR algorithm can use any optimization algorithm to compute the gradient-based update value~(Section~\ref{ssec:Gxnor_Wup}), we consider two update cases:
\begin{enumerate*}
    \item supporting general optimization algorithms, such as ADAM, and
    \item supporting the SGD algorithm.
\end{enumerate*}
Table~\ref{tab:tr_cont_sig} summarizes the circuit level signals as a function of the GXNOR variables.

\paragraph{Support of General Optimization Algorithms}
To support general optimization algorithms, the update value $\Delta W$ is computed in a peripheral circuit to the synapse array; thus, $\Delta W$ is given as an input to the array. The array columns are updated sequentially, \textit{i.e.}, a single column is updated per iteration. During this operation, the input voltages are set to $u_{1}=u_{2}=V_{up}>0$ in the active column, $u_{1}=u_{2}=-V_{dd}$ for the rest of the columns, {\color{black}and the output row interface connects the rows to ground.}
To support the stochastic update (Section~\ref{ssec:stoch_wup}), the control signals are given by
\begin{equation} 
\begin{cases}
e_{1,p}=-e_{2,p}=-sign(\Delta W) V_{dd}, & if \ \kappa\ne0 \\
e_{1,p}=e_{2,p}=V_{dd}, & else  
\end{cases}
\end{equation}
\begin{equation} 
    e_{1,n}=\begin{cases}
-sign(\Delta W)V_{dd}, & 0<t<|\nu|T_{up}\\
-V_{dd}, &|\nu|T_{up}<t<T_{up} 
\end{cases}
\end{equation}
\begin{equation} 
    e_{2,n}=\begin{cases}
sign(\Delta W)V_{dd}, & 0<t<|\nu|T_{up}\\
-V_{dd}, &|\nu|T_{up}<t<T_{up},
\end{cases}
\end{equation}
where $\Delta W$, $\nu$, and $\kappa$ are as defined in Section~\ref{sec:GXNOR}.
Hence, the MTJ is updated proportionally to $\kappa=\floor*{|\Delta W|}$ and $\nu=remainder(\Delta W/\Delta z_1)$, meaning that for a single synapse, one MTJ is updated using a pulse width of $\Delta t=\mathds{1}_{|\kappa|>0}T_{up}$, while the other is updated using a pulse width of $\Delta t=|\nu|T_{up}$. We assume that $\kappa$ and $\nu$ are inputs to the synapse array. 

\paragraph{Support of Stochastic Gradient Descent}
This update scheme is similar to the update scheme proposed in~\cite{Soudry2014}.
When the SGD algorithm is used to train the network, all the synapses in the array are updated in parallel. Therefore, in this section, we denote the array row and column indexes by $i$ and $j$, respectively.
To support SGD training, minor changes need to be made to the general update scheme. Using SGD, the update is given by the gradient value, and is equal to $\Delta W=xy^T$, where $y$ is the error propagated back to the layer, achieved using the backpropagation algorithm, and $x$ is the input. For TNNs and BNNs, the input activations are $u\in\{'-1','0','1'\}=\{-V_{up},0,V_{up}\}$ and $u\in\{'-1','1'\}=\{-V_{up},V_{up}\}$, respectively; thus, $\Delta W_{i,j}=y_iu_j$ or $\Delta W_{i,j}=0$ for $u_j=0$. In this scheme, the voltage sources retain the activation values, so $u_1=u_2=V(x)$ (whereas in the general scheme the voltage sources are set to $u_1=u_2=V_{up}$). The control signals are a function of the error $y$; so, $\kappa_{ij}=\floor*{y_i}$ and $\nu_{ij}=remainder(y_i/z_1)$ (whereas in ADAM and other optimization algorithms they are a function of the update value $\Delta W$). The control signal functionality for SGD is
\begin{equation} 
\begin{cases}
e_{i,(1,p)}=-e_{i,(2,p)}=-sign(y_i) V_{dd}, & if\ \kappa_{ij}\ne0 \\
e_{i,(1,p)}=e_{i,(2,p)}=V_{dd}, & else  
\end{cases}
\end{equation}
\begin{equation} 
    e_{i,(1,n)}=\begin{cases}
-sign(y_i)V_{dd}, & 0<t<|\nu_{ij}|T_{up}\\
-V_{dd}, &|\nu_{ij}|T_{up}<t<T_{up}
\end{cases}
\end{equation}
\begin{equation} 
    e_{i,(2,n)}=\begin{cases}
sign(y_i)V_{dd}, & 0<t<|\nu_{ij}|T_{up}\\
-V_{dd}, &|\nu_{ij}|T_{up}<t<T_{up}.
\end{cases}
\end{equation}
The functionality of the control signals remains unchanged compared to the general update scheme, except that the voltage source is selected according to $y$, and  the voltage sign and the effective update duration are set as a function of the integer $\kappa$ and the fraction $\nu$ values of $y$, respectively. Therefore, the update equation is given by
\begin{equation} 
\Delta W_{ij}=sign(y_i)sign(u_j)(\mathds{1}_{\kappa\ne0}+Bern(P_{sw}(\nu_{ij}))),
\end{equation}
where $sign(y_i)sign(u_j)=sign(\Delta W_{ij})$.

\begin{table*}[t!]
  \centering
  \caption{Weight Update -- Summary of Circuit Level Signals as a Function of GXNOR Variable Values}
  \label{tab:tr_cont_sig}
  \begin{adjustbox}{max width=1.2\textwidth,center}
    \begin{tabular}{ccccccccc}
    \toprule
       \multirow{2}{*}{\Large\textbf{Update Case}} & \multicolumn{2}{c}{\Large\textbf{Row signal}} &  \multicolumn{2}{c}{\Large\boldmath{$u_{1}=u_{2}$}} &\multirow{2}{*}{\Large\boldmath{$e_{1,n}$}} & \multirow{2}{*}{\Large\boldmath{$e_{1,p}$}} & \multirow{2}{*}{\Large\boldmath{$e_{2,n}$}} & \multirow{2}{*}{\Large\boldmath{$e_{2,p}$}}  \\ 
        & \textbf{General} & \textbf{SGD} &\textbf{General} & \textbf{SGD}  &   &  &   &  \\
    \midrule
       \Large \boldmath{$\Delta W>0$} & \Large$\Delta W>0$  & \Large$y>0$  &  \Large$V_{up}$ & \Large$u$  &  \Large$-V_{dd}$  & \Large$-V_{dd}$ & \Large
        $\begin{cases}
            V_{dd}, & 0<t<|\nu_{ij}|T_{up}\\
            -V_{dd}, &|\nu_{ij}|T_{up}<t<T_{up}.
        \end{cases}$ 
        & \Large$V_{dd}$   \\
                         
       \Large\boldmath{$\Delta W<0$} & \Large$\Delta W<0$  & \Large$y<0$  & \Large$V_{up}$  & \Large$u$  & \Large
        $\begin{cases}
            V_{dd}, & 0<t<|\nu_{ij}|T_{up}\\
            -V_{dd}, &|\nu_{ij}|T_{up}<t<T_{up}.
        \end{cases}$ 
       & \Large$V_{dd}$ & \Large$-V_{dd}$  & \Large$-V_{dd}$ \\
                      
       \Large\Large\boldmath{$\Delta W/ \kappa / \nu=0$} & \Large$\Delta W$    & \Large$y$    & \Large$V_{up}$  & \Large$u$  &  \Large$-V_{dd}$ & \Large$V_{dd}$ & \Large$-V_{dd}$ & \Large$V_{dd}$  \\
                         
    \bottomrule
    \end{tabular}
    \end{adjustbox}
\end{table*}

\subsection{Ternary Synapse Update Examples}\label{ssec:tnn_exp}
To demonstrate the proposed update scheme, two examples of synapse updates are given. Each example shows how a different weight is updated based on a calculated $\Delta W$. The control signals open the transistors according to the $\Delta W$.
\subsubsection{Example $1$: $W=-1$ and positive update value}
Figure~\ref{fig:exp1} shows the case where a synapse weight is $W=-1$ and the update value is $\Delta W=1.5$. Thus, $\kappa=1$ and $\nu=0.5$. 
Hence, $e_{1,p}=-e_{2,p}=-V_{dd}$; therefore, $P_1$ is ON and $P_2$ is OFF for time interval $T_{up}$.  $e_{1,n}=-V_{dd}$ for $T_{up}$ and $e_{2,n}$ is ON for $0.5T_{up}$, as given by
\begin{equation} 
    e_{2,n}=\begin{cases}
V_{dd} & 0<t<0.5T_{up}\\
-V_{dd} & 0.5T_{up}<t<T_{up}.
\end{cases}
\end{equation}
Hence, the probability of $R_1$ and $R_2$ switching is $P_{sw,1}\approx1$ and $P_{sw,2}=P(0.5T_{up})$, respectively.
In this example, the synapse weight will be updated from $W=-1$ to $W=0$ with a probability of
\begin{equation} 
\begin{array}{l}
P_{-1 \rightarrow 0}=P_{-1 \rightarrow 0_w}+P_{-1 \rightarrow 0_s}=
P_{sw,1}(1-P_{sw,2})+(1-P_{sw,1})(1-P_{sw,2}) 
\\\approx (1-P_{sw,2}),    
\end{array}
\end{equation}
and can switch to $1$ with a probability of
\begin{equation} 
P_{-1 \rightarrow 1}=P_{sw,1}P_{sw,2} \approx P_{sw,2}.
\end{equation} 
Note that when $W=-1$, $\{R_1,R_2\}=\{R_{off},R_{on}\}$. Thus, if $\Delta W<0$, due to the current that flows from $R_2$ to $R_1$, both MTJs cannot switch and the state will remain unchanged.
\subsubsection{Example $2$: $W=0$ and negative update value}
\begin{figure}[!t]
\centering
        \begin{subfigure}{.35\textwidth}
        \centering
                \includegraphics[trim={6cm 0cm 5.5cm 0cm},clip,width=1\linewidth]{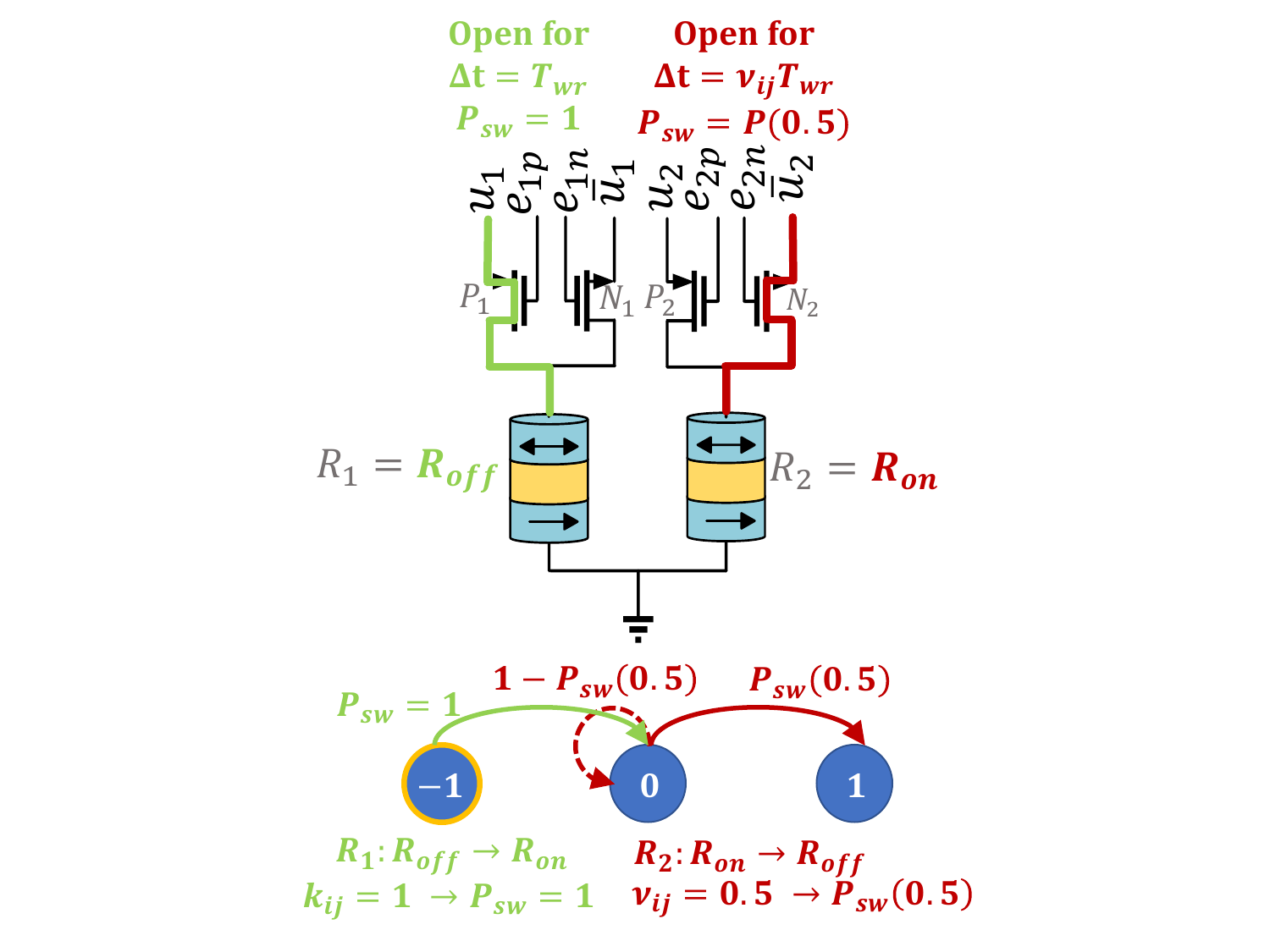}
                \caption{Synapse update where\\ $W=-1$ and $\Delta W=1.5$}
                \label{fig:exp1}
        \end{subfigure}\hspace{0.6ex}
        \begin{subfigure}{.35\textwidth}
        \centering
                \includegraphics[trim={6cm 0cm 6cm 0cm},clip,width=1\linewidth]{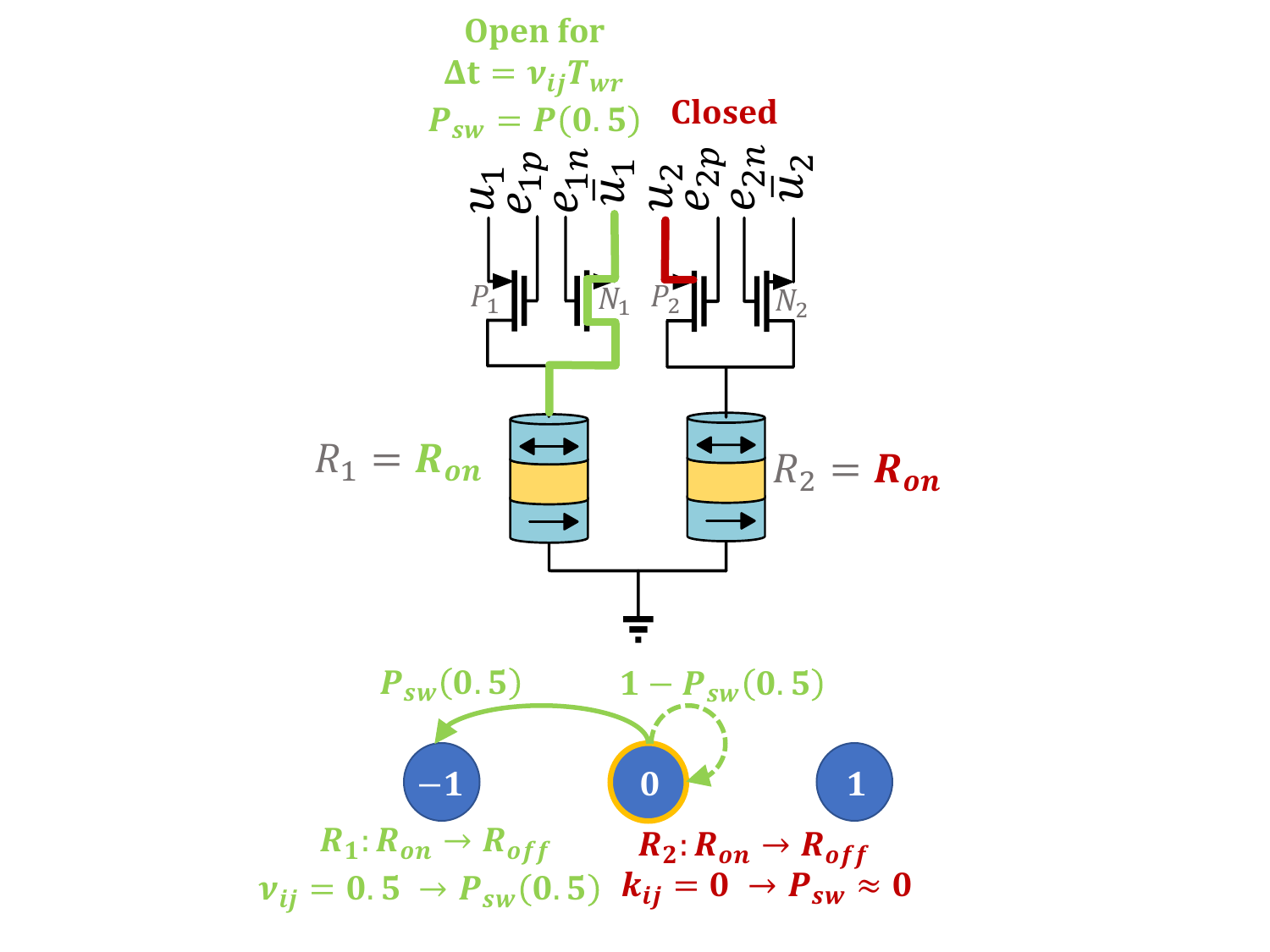}
                \caption{Synapse update where \\$W=0_w$ and $\Delta W=-0.5$}
                \label{fig:exp2}
        \end{subfigure}
        \caption{Examples of synapse updates. Blue circles represent the logic state of the weight, where the initial state is marked by an orange outline.}
        \label{fig:Update_examples}
\end{figure}

Figure~\ref{fig:exp2} shows the case where a synapse weight is $W = 0_w$ and the update value is $\Delta W = -0.5$. Thus, $\kappa=0$ and $\nu=-0.5$. 
Consequently, $e_{1,p}=e_{2,p}=V_{dd}$, so both $P_1$ and $P_2$ are closed for $T_{up}$. $e_{2,n}=-V_{dd}$ for $T_{up}$ and the transistor connected to $e_{1,n}$ is open for $0.5T_{up}$, as given by
\begin{equation} 
    e_{1,n}=\begin{cases}
V_{dd} & 0<t<0.5T_{up}\\
-V_{dd} & 0.5T_{up}<t<T_{up}.
\end{cases}
\end{equation}
Therefore, only $R_1$ can switch with a probability of $P_{sw,1}=P(0.5T_{up})$.
In this example, the synapse weight is updated from $W=0_w$ to $W=-1$, with a probability $P=P_{sw,1}$. Although, theoretically, no current should flow through $R_2$, it might switch from $R_{on}$ to $R_{off}$ due to leakage currents with probability $P_{sw,2}<<1$.

\subsection{Feedforward and Backpropagation}
\label{ssec:TNN_FFnBP}
TNN training requires the circuits to support the feedforward and backpropagation stages~\cite{tz_mom}. The feedforward stage requires to compute matrix-vector or matrix-matrix multiplication; in TNNs, the multiplication is replaced with the gated XNOR (GXNOR) operation. The GXNOR logic outputs zero if one of the inputs is zero; otherwise, it outputs the XNOR operation between the inputs.

In this section, we first explain how our synapse performs the GXNOR operation and then how the synapse array is used to perform the near-memory quantized matrix-vector multiplication.
When supporting training, the feedforward stage is followed by computation of the error of each layer, known as the backpropagation stage. This stage requires computation of the matrix-vector multiplication between the transposed weight matrix ($W^T$) and the layer error vector ($y$). In the following sections, we denote the matrix-vector multiplication of $W^Ty$ as ``backpropagation".

\subsubsection{Gated XNOR}
To perform the GXNOR logic operation between the synapse and activation values~\cite{GXNOR}, we denote the input neuron values as the voltage sources. Accordingly, $u=V(a)$ is the voltage representing input value $a$. The logic values of the input neuron $a\in\{-1,0,1\}$ are represented by $u\in\{-V_{rd},0,V_{rd}\}$. $V_{rd}$ is set to guarantee the low current regime of an MTJ, so the switching probability is negligible. During this operation, $u_1=u$, $\Bar{u}_2=-u$, $\{e_{(1,n)},e_{(1,p)},e_{(2,n)},e_{(2,p)}\} = \{-V_{dd},-V_{dd},V_{dd},V_{dd}\}$ and the synapse output node is grounded. The result is given by the output current sign, 
\begin{equation} 
I_{out}=(G_1-G_2)u,
\end{equation}
where $G_{\{1,2\}}$ is the conductance of each MTJ device, respectively.
As listed in Table~\ref{tab:syn_state}, the polarity of $I_{out}$ depends on the input voltage and the synapse weight. If $u=0$ or $W=\{0_w,0_s\}$, the output current is $I_{out}\approx0$. If the weight and input have the same polarity, then $sign(I_{out})=1$, else $sign(I_{out})=-1$.

\subsubsection{Feedforward}
The quantized feedforward operation is given by
\begin{equation} 
O_m = \sum_{n=1}^{N}GXNOR(W_{mn},x_n), \forall m\in[1,M],
\end{equation}
where $O_m$ is the result of row $m$. 
During this operation, each column voltage is mapped to the corresponding input activation ($u_n=V(a_n)$, $\forall n \in [1,N]$), {\color{black}the output row interface connects the rows to ground. Thus, each synapse computes the GXNOR operation between its input and stored weight and} the output currents from all synapses are summed based on KCL. Thus, the current through row $i$ is
\begin{equation}
\begin{aligned}
 I_{row,i}=&\sum_{j=1}^{N}(G_{ij,R_1}-G_{ij,R_2})u_j=       &\frac{R_{off}-R_{on}}{R_{off}R_{on}}\big(N_{+1,i}-N_{-1,i}\big)V_{rd},
\end{aligned} 
\end{equation}
where $G_{ij,R_{\{1,2\}}}$ is the conductivity of each MTJ, $N$ is the number of synapses per row, $N_{+1,i}$ is the total number of positive products in row $i$, and $N_{-1,i}$ is the total number of negative products in row $i$.

\subsubsection{Backpropagation}
To train the TNN, backpropagation of the error must be performed. Therefore, the synapse array also supports the matrix-vector multiplication $W^Ty$. Rather than storing $W^T$ in a dedicated array, we reuse the same array, which stores $W$, similarly to \cite{Soudry2014}. During this operation, the output row interface is used as an input and the output is given by the current measured in the columns.
Due to the synapse structure, the current is separated into two columns, as shown in Figure~\ref{fig:inv_rd}. Therefore, the operation result is given by the current difference
\begin{equation} 
 I_{col,j}=\sum_{i=1}^{M}(I_{ij,R_1}-I_{ij,R_2}) = \sum_{i=1}^{M}(G_{ij,R_1}-G_{ij,R_2})y_i,
\end{equation}
where $y_i$ is the layer's error. The current through each column pair is converted to voltage, and the result is computed using a voltage comparator. 

\section{Evaluation and Design Considerations}\label{sec:eval}
This section presents the performance evaluation of the MTJ-based QNN training. 
The functionality, power and area of the synapse circuit and array were evaluated in Cadence Virtuoso and used for the training simulations. The MTJ-based design of the GXNOR algorithm (MTJ-GXNOR) is compared to a software implementation of the algorithm (GXNOR in our terminology).

\subsection{Methodology}
Our proposed circuit is a hardware implementation of the GXNOR framework used to train QNNs~\cite{GXNOR}. We evaluate our design using four metrics:
\begin{enumerate}
    \item \textit{Circuit Evaluation (Section~\ref{subsec:circ_eval})}. We validated the circuit operations needed to support the MTJ-GXNOR framework. Our circuit needs to support a stochastic update, the GXNOR, and backpropagation operations. The MTJ switching operation was evaluated by running Monte Carlo simulation of the MTJ transient response. We evaluated the GXNOR and backpropagation operations using SPICE simulations. 
    \item \textit{Training Simulation (Section~\ref{subsec:train_eval})}. To validate that our proposed synapse can be used to train QNNs and reach comparable results to the GXNOR algorithms and other state-of-the-art QNN frameworks~\cite{GXNOR,Courbariaux2016,Courbariau2015}, we simulated MTJ-GXNOR training in PyTorch with the hardware circuit parameters extracted from the circuit evaluation.
    \item \textit{MTJ-GXNOR Sensitivity to Process Variation (Section~\ref{ssec:proc_var})}. The MTJ-GXNOR training performance is influenced by the device variation and environmental changes. Hence, we evaluated the sensitivity of the MTJ-GXNOR test accuracy considering process and environmental variations.
    \item \textit{Hardware Performance Evaluation (Section~\ref{subsec:perf_eval})}. Our design can be integrated into different architectures, each leading to a different performance. Here, we report on the performance of our basic cells -- the hardware synapse and synapse array. We also consider a simple system test case comparable with previous solutions.
\end{enumerate}  

\subsection{Circuit Evaluation}
\label{subsec:circ_eval}
The synapse circuit was designed and evaluated in Cadence Virtuoso for the GlobalFoundries $28$nm FD-SOI process. The MTJ device is based on device C from~\cite{Analitic_MTJ} and its parameters are listed in Table~\ref{tab:MTJ}. To achieve higher switching probability, the magnetization saturation ($\mu_0M_s$) was changed according to~\cite{Ms_scale}.

The read voltage, $V_{rd}$, was set to guarantee a low-current regime and negligible switching probability for the feedforward and backpropagation operations.
Likewise, the update voltage, $V_{up}$, was set to guarantee a high-current regime. The update time period was set to match $P_{sw}\big(T_{up}\big)\approx1$.

\subsubsection{Circuit Schematic Model}
The transistors and interconnect affect the circuit's functionality and performance. Therefore, we adopted a circuit model that considers the parasitic resistance and capacitance of wires and transistors. {\color{black}the model considers the location of the synapse. An illustration of the schematic circuit model appears in the supplementary material (Section $2$). Using the schematic circuit model, and SPICE simulations, we evaluate the circuit array and operations.}  
We considered the corner cases, \textit{i.e.}, the synapses located at the four corners of the synapse array, to evaluate the worst-case scenario for the effect of the wires and transistors on operation results, latency, and power consumption. 
For all circuit simulations, we considered the worst case, where the wire resistance and capacitance are the most significant, $i.e.$, for an array of size $M\times N$, the synapse located at [M,1].

\subsubsection{MTJ Switching Simulation}
To evaluate the transition in the resistance of the MTJ and its impact  on the operation of the synapse, we performed a Monte Carlo simulation of the MTJ operation. The simulation numerically solves the Landau\textendash Lifshitz\textendash Gilbert (LLG) ~\cite{Gilbert2004,Vincent2015} differential equation (assuming the MTJ is a single magnetic domain) with the addition of a stochastic term for the thermal fluctuations~\cite{GarcuaPalacios1998} and Slonczewski's STT term~\cite{Slonczewski1996}. For each iteration of the Monte Carlo simulation, a different random sequence was introduced to the LLG equation and the resulting MTJ resistance trace was retrieved. The equation was solved using a standard midpoint scheme~\cite{Aquino2006} and was interpreted in the sense of Stratonovich, assuming no external magnetic field~\cite{Diao2007} and a voltage pulse waveform. The resistance of the MTJ was taken as $R_{on}\frac{1+P^2}{1+P^2cos\theta}$~\cite{Slonczewski1989}, where $\theta$ is the angle between magnetization moments of the free and fixed layers and $P$ is the spin polarization of the current.
To approximate the time-variation resistance of an MTJ during the switch between states, all the traces from the Monte Carlo simulation were aligned using the first time that the resistance of the MTJ reached $\frac{R_{on}+R_{off}}{2}$. After the alignment, a mean trace was extracted and used for the fit. This fit was used as the time-variation resistance when the MTJ made a state switch.

\begin{table}[t!]
    \centering
    \caption{Circuit Parameters}
    \begin{tabular}{c c c c }
    \hline
        \multicolumn{4}{c}{\textbf{MTJ, device C~\cite{Analitic_MTJ}}} \\
                    \hline
        \textbf{Parameter} & \textbf{Value} & \textbf{Parameter} & \textbf{Value} \\
                    \hline                 
        $a$ [nm] & $50$  & Temp. [K] & 300 \\
                    \hline
        $b$ [nm] & $20$  & $R_{on} [\Omega]$ & 1500 \\
                    \hline
        $t_f$ [nm] & $2.0$ & $R_{off} [\Omega]$ & 2500 \\
                    \hline        
        $\mu_0M_s$ [T]\text{\footnotemark[1]} & $0.5$  & $\alpha$ & 0.01 \\
                    \hline
        $I_{c_o}$ [$\mu$ A] & $157$  & $\theta_0$ & 0.345 \\
                    \hline \hline  
        \multicolumn{4}{c}{\textbf{CMOS}} \\
                    \hline
        \textbf{Parameter} & \textbf{Value} & \textbf{Parameter} & \textbf{Value} \\
                    \hline                          
        $V_{DD}$ [V] & $1$  &W/L$_{PMOS}$& $33$ \\
                    \hline
        $V_{SS}$ [V] & $-1$ &W/L$_{NMOS}$& $20$ \\
                    \hline
        $V_{up}$ [V] & $1$ & $T_{up}$ [ns] & 2 \\
                    \hline        
        $V_{rd}$ [V] & $0.1$ & $T_{rd}$ [ns] & 0.5 \\
                    \hline 
        \multicolumn{4}{l}{\text{\footnotemark[1] \footnotesize To achieve higher switching probability, the value}}\\
        \multicolumn{4}{l}{\text{\footnotesize of $\mu_0M_s$ was changed according to~\cite{Ms_scale}.}}\\
    \end{tabular}
    \label{tab:MTJ}
\end{table}

\subsubsection{GXNOR Operation}
\begin{figure*}[t!]
    \centering
    \begin{adjustbox}{max width=1.4\textwidth,center}
    \includegraphics[trim={0cm 11.6cm 4cm 0cm},clip,width=1.2\textwidth]{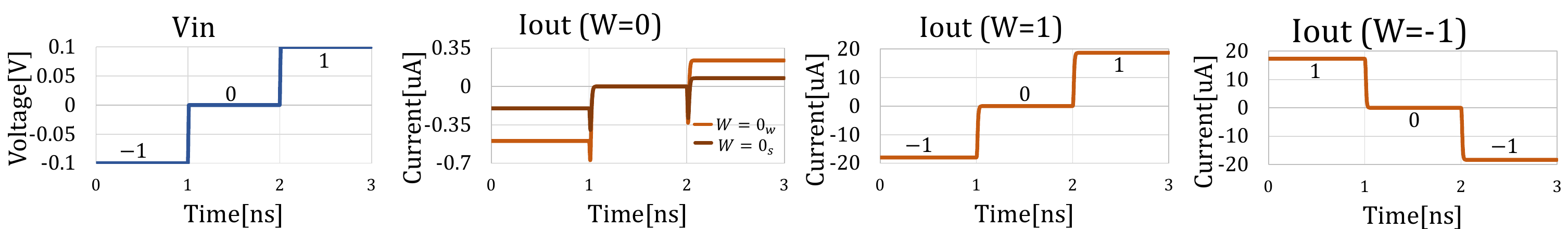}
    \end{adjustbox}
    \caption{GXNOR operation between the input voltage $V_{in}\in \{-1,0,1\}=\{-V_{rd},0,V_{rd}\}$ and the weight value ($W$ in the figure). During the GXNOR operation (read operation), $V_{rd}$ is $0.1V$ to guarantee a low-current domain and low switching probability. For $V_{in}\ne 0$ and $W=0_{w/s}$, the output current is not zero. This is a source for error when the GXNOR results are summed to compute the activation value. Limiting the dimensions of the synapse array can mitigate this effect.}
    \label{fig:GXNOR}
\end{figure*}

The GXNOR operation for a single synapse is shown in Figure~\ref{fig:GXNOR}. When either the activation (input) or the weight ($W$) is zero, the output current is one order of magnitude lower than in the other cases. 

\subsection{MTJ-GXNOR Training Simulation}
\label{subsec:train_eval}
To evaluate the training performance of our solution, we determined the test accuracy, and compare it to the original GXNOR algorithm implemented in software and to other state-of-the-art frameworks. We denote our results as `MTJ-GXNOR' and the ideal GXNOR algorithm as `GXNOR'. We tested the quantized networks over the MNIST, SVHN and CIFAR10 datasets~\cite{MNIST,SVHN,CIFAR10}. 
The following three quantization resolutions were simulated in PyTorch:
\begin{enumerate*}
    \item a full ternary network (`MTJ-GXNOR TNN'),
    \item a full binary network (`MTJ-GXNOR BNN'), and
    \item a network with ternary weight and binary activations (`MTJ-GXNOR Bin-Activation').
\end{enumerate*}
For the MNIST and SVHN dataset we trained the same convolution neural networks described in~\cite{GXNOR}.
The network architecture is ``32CONV5-MP2-64CONV5-MP2-512FC” for the MNIST dataset, and ``2$\times$(128CONV3)-MP2-2$\times$(256CONV3)-MP2-2$\times$(512CONV3)-MP2-1024FC"
for the SVHN dataset, where CONV, MP and FC are the convolution layer, maxpool layer and fully connect layer, respectively. For the CIFAR$10$, we trained the VGG$16$ network~\cite{VGG}. We trained the networks using ADAM optimization algorithms with batch sizes of $100$ for the MNIST, $1000$ for SVHN and $750$ for CIFAR$10$. Table~\ref{tab:Training} lists the test accuracy of MTJ-GXNOR compared to GXNOR and other state-of-the-art algorithms (ideal software implementations). For the MNIST and SVHN datasets our solution achieved accuracy similar to that of the state-of-the-art algorithms, implemented in software. For the CIFAR$10$, the MTJ-GXNOR reached accuracy comparable to that of GXNOR, but lower compared to the other algorithms. Notwithstanding, considering the application and the hardware performance improvement, some accuracy degradation might be acceptable.

Although the BNN~\cite{Courbariaux2016} and BWN~\cite{Courbariau2015} frameworks achieve better results compared to the GXNOR BNN~\cite{GXNOR}, they retain the full-precision weights during the training phase, which increases the frequency of memory accesses and requires support of full-precision arithmetic. Hence, their potential hardware implementation will be much less efficient than GXNOR. The MTJ-GXNOR TNN results are similar to the results of the GXNOR training, showing less than $1\%$ accuracy degradation.  Compared to GXNOR BNN, the MTJ-GXNOR BNN results led to less than $1\%$ accuracy degradation for the MNIST dataset, but $2.4\%$ degradation for the SVHN dataset. The test accuracy of the MTJ-GXNOR Bin-Activation, which used ternary weights and binary activations, is closer to that of GXNOR TNN. 

\begin{table}[t!]
    \centering
    \caption{Accuracy of State-of-the-art Algorithms}
    \begin{adjustbox}{max width=0.6\textwidth,center}  
    \begin{tabular}{c c c c}
    \hline
        \multirow{2}{*}{\textbf{Methods}}     & \multicolumn{2}{c}{\textbf{Datasets}} \\\cline{2-4}
        \rule{0pt}{3ex}            & \textbf{MNIST}    &  \textbf{SVHN}  &  \textbf{CIFAR10}     \\
                    \hline       
        BNNs~\cite{Courbariaux2016} & $98.6\%$ & $97.20\%$ & $89.85\%$\\
                    \hline
        TWNs~\cite{Li2016} & $99.35\%$ & N.A  & $92.56\%$\\
                    \hline
        BWNs~\cite{Courbariau2015} & $98.82\%$ & $97.70\%$ & $91.73\%$ \\
                    \hline
        BWNs~\cite{Li2016} & $99.05\%$ & N.A & $90.18\%$\\
                    \hline        
        GXNOR TNN~\cite{GXNOR} & $99.32\%$ & $94.12\%$ & $83.51\%$ \\
                    \hline    
        GXNOR BNN~\cite{GXNOR} & $98.54\%$ & $91.68\%$ & N.A\\
                    \hline                     
    \makecell{\textbf{MTJ-GXNOR }\\\textbf{TNN}} & $\mathbf{98.61}$\textbf{\%}  & $\mathbf{93.99}$\textbf{\%} & $\mathbf{83.02}$\textbf{\%}\\
                        \hline  
    \makecell{\textbf{MTJ-GXNOR} \\ \textbf{Bin-Activation}} & $\mathbf{98.6}$\textbf{\%}  & $\mathbf{93.62}$\textbf{\%} & --\\
                        \hline 
    \makecell{\textbf{MTJ-GXNOR}\\ \textbf{BNN Full}} & $\mathbf{97.84}$\textbf{\%}  & $\mathbf{89.46}$\textbf{\%} & -- \\
                        \hline                             
    \end{tabular}
    \end{adjustbox}
    \label{tab:Training}
\end{table}

\subsection{Training Performance Sensitivity to Process Variation}
\label{ssec:proc_var}
Device variation and environmental changes may affect the performance of the proposed circuits, including their training performance. In this section, we evaluate the sensitivity of the TNN training performance to process variation.

\subsubsection{Resistance Variation and $\theta$ Distribution Variation}
\label{sss:R_theta_var}
Two cases of process variation were considered:
\begin{enumerate*}
  \item resistance variation and 
  \item variation in the distribution of $\theta$.
\end{enumerate*}
These variations may lead to a different switching probability for each MTJ device.
To evaluate the sensitivity of the training to the device-to-device variation, we simulated the MNIST-architecture training with variations in the resistance and $\theta$ distributions.
Several Gaussian variabilities were examined with different relative standard deviations (RSD), where the mean values are shown in Table~\ref{tab:MTJ}. Table~\ref{tab:TrainingVar} lists the training accuracy for resistance variation and variation in $\theta_0$. Typically, resistance RSD is approximately $5\%$~\cite{Vincent2015}, while our simulations show that the training accuracy is robust to resistance variation even for higher RSD values (\textit{e.g.}, only $0.46\%$ accuracy degradation for RSD$=30\%$).

The training accuracy is more sensitive to variations in $\theta_0$. Nevertheless, high standard deviation of $\theta$ values resulted in better test accuracy. 
To further evaluate the test accuracy dependency on $\theta_0$, we simulated training for different $\theta_0$ values. Table~\ref{tab:VarTheat} lists the training results. Larger $\theta_0$ values, which correspond to higher switching probability, resulted in better test accuracy. Thus, we conclude that the performance of the MTJ-GXNOR algorithm improves for higher switching probabilities, which corresponds to larger $\theta_0$ values.
\subsubsection{Sensitivity to Voltage Non-Ideality}
Since the weight update probability is a function of the voltage drop across the MTJ device ($V_{up}$), it is sensitive to voltage source variation.
Higher voltage leads to higher current. We tested training with $V_{up}$ in the range of $[0.5V,2.5V]$. Our results show that the test accuracy improves when increasing $V_{up}$.
The voltage magnitude can, therefore, be used to control the stochastic switching process and to improve the network training performance when using an MTJ device with low $\theta_0$ variance.
In our setup, this effect is bounded and diminished when $V_{up}$ exceeded $1.1V$ and only marginally improves the test accuracy; hence, in this work, we set $V_{up}=1V$ to constrain the power consumption of our design.

\begin{table}[t]
    \centering
    \caption{Test Accuracy vs. Process Variation for MNIST}
    \lineup
    \begin{adjustbox}{max width=0.4\textwidth,center}    
    \begin{tabular}{c c c}
    \toprule
        \textbf{RSD} & \textbf{Resistance} & \textbf{$\theta_0$} \\
                     & \textbf{Variation} & \textbf{Variation} \\
        \midrule    
        $0\%$  & $98.61\%$ & $98.61\%$ \\
        $1\%$  & $98.13\%$ & $97.98\%$ \\
        $5\%$  & $98.13\%$ & $97.92\%$\\      
        $10\%$ & $98.1\%$ & $97.98\%$\\
        $30\%$ & $98.15\%$ & $98.05\%$\\
        $35\%$ & $97.94\%$ & $98.05\%$\\
    \bottomrule
    \end{tabular}
    \end{adjustbox}
    \label{tab:TrainingVar}
\end{table}

\begin{table}[t]
    \centering
    \caption{Test Accuracy vs. $\theta_0$}
    \begin{adjustbox}{max width=0.35\textwidth,center}
    \begin{tabular}{c c}
        \toprule
      $\mathbf{\theta_0}$\textbf{[rad]} & \textbf{Test Accu.} \\
        \midrule        
        $0.0913$  & $94.28\%$ \\
        $0.1141$  & $94.98\%$ \\
        $0.2739$  & $97.39\%$\\
        $0.345$   & $98.61\%$\\
         \bottomrule
    \end{tabular}
    \end{adjustbox}
    \label{tab:VarTheat}
\end{table}

\subsubsection{Sensitivity to Temperature}
The MTJ dependency on temperature has several aspects. First, the switching behavior depends on the ambient temperature (\ref{eq:T_swHC}). For higher temperatures, the mean switching time $\tau$ is shorter~\cite{Temp1}. Second, higher temperatures lower $R_{off}$. The resistance of $R_{on}$, however, has a much weaker temperature dependency and it is nearly constant~\cite{Temp2}. The transistors are also influenced by the temperature. For high temperatures, the current drivability of the MOS transistors is degraded since the electron mobility is lower.
Hence, ambient temperature affects the switching probability by lowering $R_{off}$ and degrading the CMOS current drivability.
Furthermore, the initial magnetization angle, $\theta$, depends on the temperature by the normal distribution $\theta \sim \mathcal{N}(0,\theta_0)$, where the standard deviation is $\theta_0=\sqrt{k_BT/(\mu_0H_kM_sV)}$. Hence, $\theta_0$ increases for higher temperatures. 
\par
\begin{figure}[t!]
    \centering
    \includegraphics[trim={0cm 1cm 0cm 0cm},width=0.5\columnwidth]{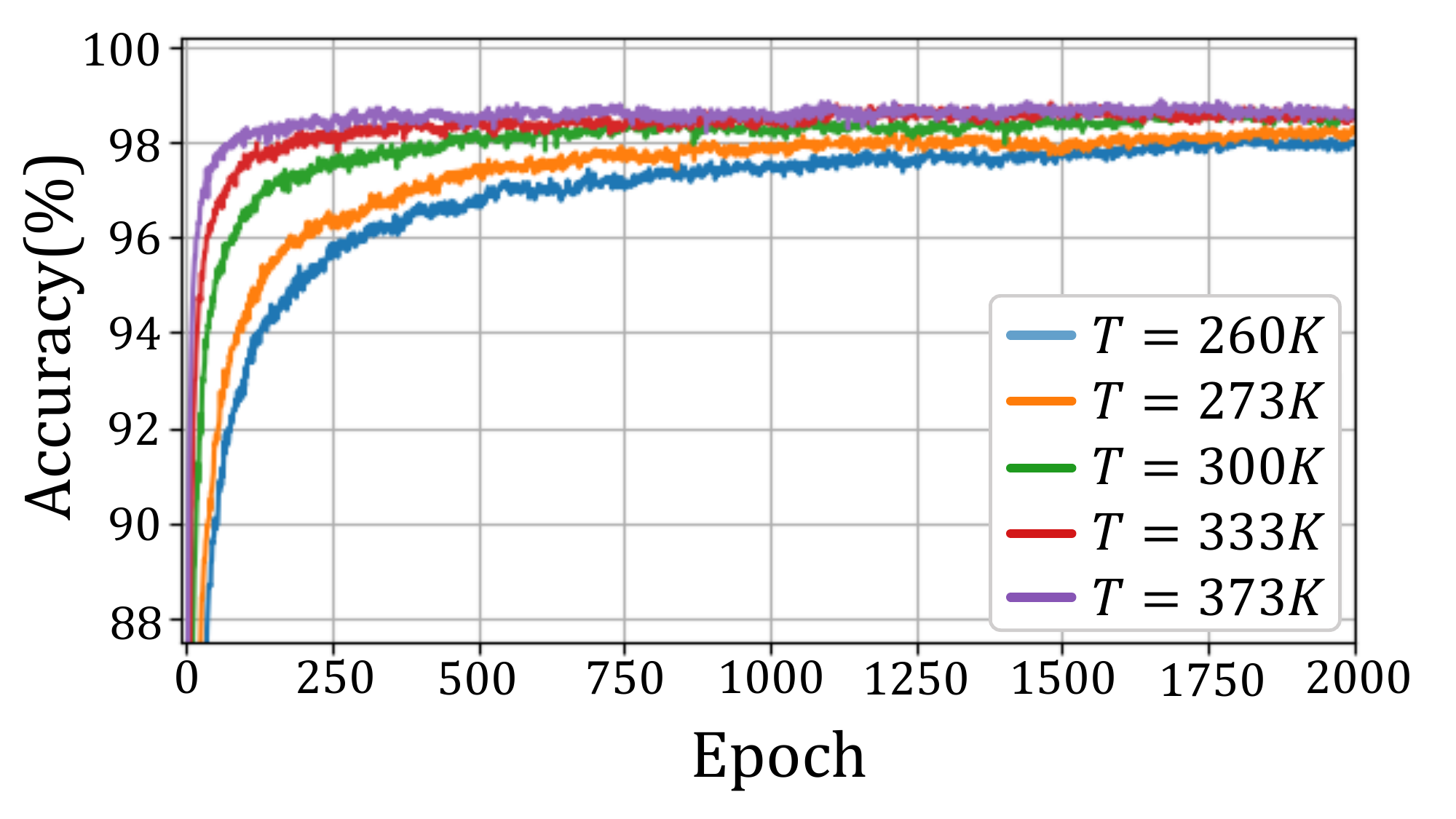}
    \caption{Test accuracy during the training phase for temperature range $[273K,373K]$. Increasing the temperature leads to higher $\theta_0$ variance, thereby increasing the randomness of the MTJ switching time. Therefore, higher temperature leads to faster convergence.}
    \label{fig:T_rst}
\end{figure}

 As shown in Section~\ref{sss:R_theta_var} and Table~\ref{tab:TrainingVar}, the training performance is influenced by the value of $\theta$, {\color{black} when in this work we do not include the effect of the transistors}. Thus, to evaluate the sensitivity of the MTJ-GXNOR training to ambient temperature, we focused on $\theta_0$. 
We simulated MTJ-GXNOR training with different temperatures in the range [$260K,373K$], with the associated resistance based on~\cite{Temp1,Temp2}.
Table~\ref{tab:Temp_Training} lists the test accuracy obtained for different temperatures. Although better accuracy was obtained for higher temperatures, the improvement was less than $1\%$. The minor variations in accuracy imply that the test accuracy is agnostic to the temperature.
Figure~\ref{fig:T_rst} shows the test accuracy over the training phase for the MNIST network. Higher temperatures increased the convergence rate of the network, while the network converged to similar test accuracy for all the temperatures in the examined range.

\begin{table*}[t!]
    \centering
    \caption{Temperature Effect on Test Accuracy for MNIST}
    \begin{adjustbox}{max width=0.6\textwidth,center}
    \begin{tabular}{c c c c c c}
        \toprule    
       \textbf{T[K]} & $\mathbf{260}$ & $\mathbf{273}$ & $\mathbf{300}$ & $\mathbf{333}$ & $\mathbf{373}$  \\
        \midrule
        $R_{off}[\Omega]$ & $2780$ & $2690$ & $2500$ & $2270$ & $2000$  \\
                    
        $\theta_0[rad]$ & $0.3187$ & $0.3266$ & $0.345$ & $0.3617$ & $0.3827$  \\
                                          
        \textbf{Test Accuracy(\%)} & $\mathbf{98.14}$ & $\mathbf{98.32}$ & $\mathbf{98.66}$ & $\mathbf{98.82}$ & $\mathbf{98.88}$  \\
       \bottomrule             
    \end{tabular}
    \end{adjustbox}
    \label{tab:Temp_Training}
\end{table*}

\begin{table}[t!]
    \centering
    \caption{Area and Power}
    \begin{adjustbox}{max width=0.6\textwidth,center}
    \begin{tabular}{c c c c c }
    \toprule
      \multirow{ 2}{*}{\textbf{Cell}}& \multirow{ 2}{*}{\textbf{Area}} & \multicolumn{2}{c}{\textbf{Power}} \\
        & & \textbf{XNOR+Sum} &\textbf{Update} \\
        \midrule                   
        Single synapse & $3.63 \mu m^2$ & $1.89\mu W$&$2.72\mu W$ \\
         
        $64\times64$ Syn. array & $0.015 mm^2$ & $7.31mW$ & $1.64mW$ \\
         
        $128\times128$ Syn. array & $0.059 mm^2$ & $28.5mW$ & $3.25mW$ \\
        \bottomrule           
    \end{tabular}
    \end{adjustbox}
    \label{tab:perf}
\end{table}

\subsection{Performance Evaluation}
\label{subsec:perf_eval}
QNNs aim to reduce the computation and memory-capacity requirements of DNNs; therefore, in this section, we evaluate the potential performance benefits of our solution. 
The overall performance is highly dependent on the exact system structure and functionality. For example, our solution can be integrated into a fully analog or digital architecture and can support different general optimization algorithms, including SGD, as in GXNOR~\cite{GXNOR}. Each configuration will produce different performance and should be compared to a similar configuration. 
First, we evaluate the performance of our circuit when it is not connected to the peripheral circuit. Then, we consider a simple test system to evaluate the potential of our solution with its associated peripheral circuits and supporting units.

\subsubsection{TNN Power and Area}
The power consumption and area were evaluated for a single synapse and different synapse arrays simulated in Cadence Virtuoso, including the interconnect parasitic. The results are given in Table~\ref{tab:perf}. During the read operation, all the synapses are read in parallel; therefore, the feedforward power is higher than the write power, when each column is updated separately. 

\subsubsection{System Performance (Test Case)} 
\begin{figure}[t!]
    \centering
    \includegraphics[width=0.7\columnwidth]{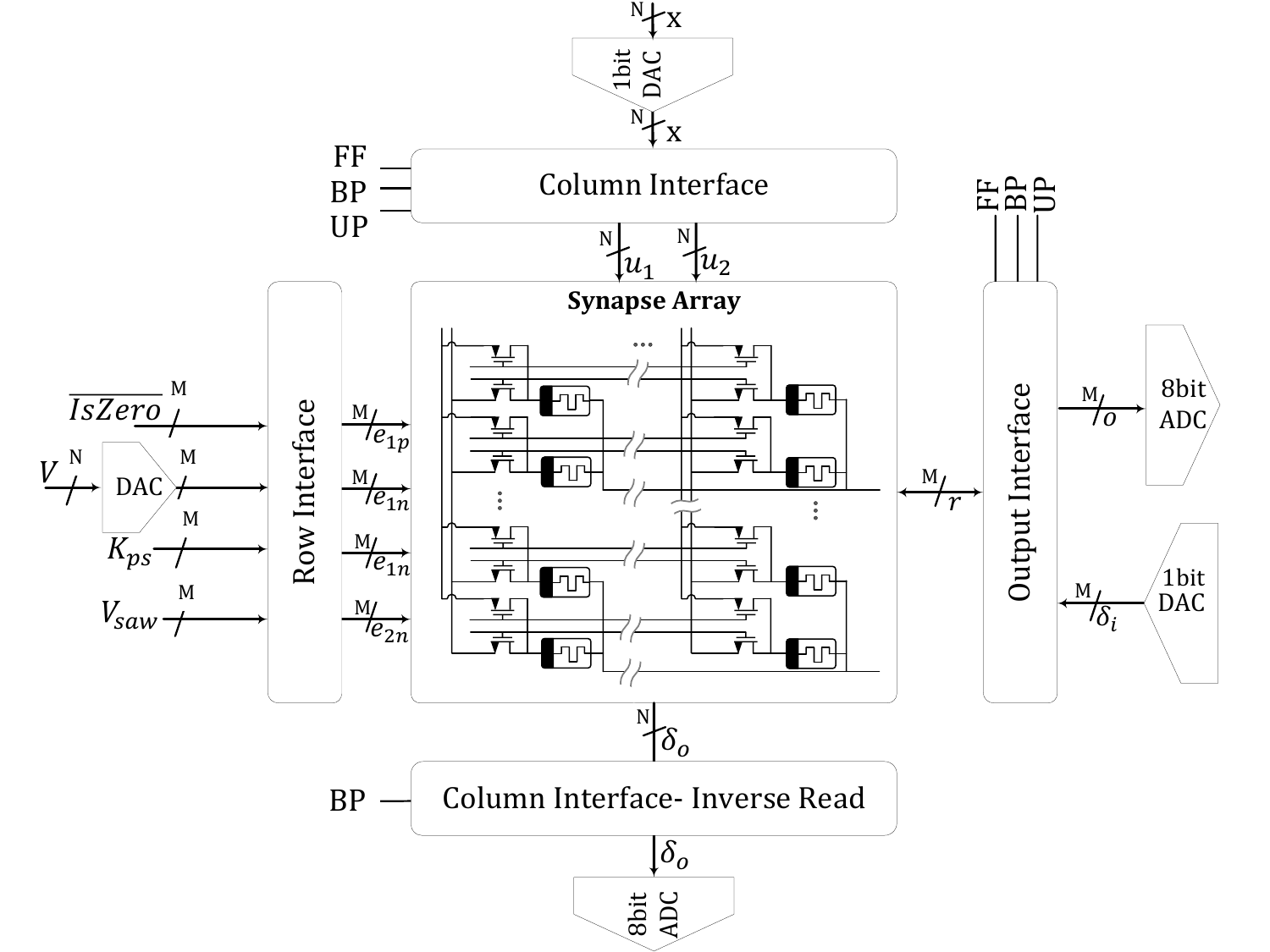}
    \caption{System configuration -- test case. A $127\times 127$ synapse array is connected to the peripheral circuits through the row and column interfaces. The FF, BP, and UP control signals configure the interfaces to support the feedforward, backpropagation and weight update, respectively. $\kappa$ and $\nu$ are inputs to the array. The $IsZero$ signal indicates if $\kappa$ is zero.}
    \label{fig:TestSys}
\end{figure}

To evaluate the performance of the synapse array when integrating our design in a full system, we consider the following setup, illustrated in Figure~\ref{fig:TestSys}.
\begin{enumerate}
    \item The synapse array stores the ternary weights. The array can perform the GXNOR, backpropagation, and GXNOR operation, as described in Section~\ref{sec:MTJ_TNN_syn}. 
    \item A $127\times 127$ synapse array. 
    This size is broadly accepted as mitigating the parasitic effects on the circuit performance {\color{black}that is limited by the ADC resolution~\cite{ISAAC,PipeLayer,Edouard}.}
    Since the size of a single DNN layer is larger than the array size, the layer will be divided between different arrays and the partial results are accumulated. To support a multi-array per layer, the binary activation is done after accumulating the results from the array. Thus, this system requires conversion of the partial results ($I_{out}$), from each array, to digital using analog-to-digital converters (ADC). 
    \item For the feedforward, $1$-bit DACs are used to support the inputs per column, and $8$-bit ADCs are used to convert the row current to digital outputs. For a $127\times 127$ array, the output of each row is an integer value in the range $[-127,127]$; thus an $8$-bit ADC is sufficient. 
    Furthermore, due to the high energy consumption of the ADC, we use only eight ADCs, which are shared among the $127$ output rows~\cite{ISAAC}. Accordingly, the overall latency to produce $127$ sum results is $8$ns.
    \item For the backpropagation, we consider the bit-streaming method with $8$-bit precision as suggested in~\cite{ISAAC}; thus, we used a $1$-bit DAC in the row interface and an $8$-bit ADC in the column outputs.
    \item To generate the control signals, an $8$-bit DAC, and voltage comparators are needed. 
    To generate the sawtooth signal, we use the circuit from~\cite{Eyal_Sergey}.
    \item The system supports an in-situ SGD algorithm. Therefore, no additional circuit is needed to compute the update values. The columns are updated sequentially.
\end{enumerate}
Table~\ref{tab:periph} lists the power of the additional peripheral circuits.
The energy efficiency of each stage for this setup is listed and compared to previews works in Table~\ref{tab:Eeff}.
The power consumption of the data converters significantly limits the overall performance. 
\begin{table}[t!]
    \centering
    \caption{Test Case Model}
    \begin{tabular}{c c c c}
    \toprule
         \textbf{Component}     & \textbf{Number}       & \textbf{Power} $\mathbf{[mW]}$ \\
        \midrule
        ADC $8$-bit~\cite{ISAAC}    & $16$          & $16$        \\
           
        DAC $8$-bit~\cite{DAC}     & $2\times127$ & $5.47$      \\        
           
        DAC $1$-bit~\cite{ISAAC}   & $2\times127$ & $0.5$         \\
        Voltage Comparator~\cite{VComp} & $ 127$ & $0.455$         \\
        \bottomrule          
    \end{tabular}
    \label{tab:periph}
\end{table}
\begin{table}[t!]
    \centering
    \caption{Energy Efficiency of the Synaptic Array.}
    \begin{adjustbox}{max width=1.4\textwidth,center}
    \begin{tabular}{c c c c c c c c c}
    \toprule
         \textbf{Paper}   & {\color{black}\textbf{Tech.}}   & {\color{black}\textbf{FF}}  & {\color{black}\textbf{BP}}  & {\color{black}\textbf{WU}} & {\color{black}\textbf{Delay}}   & {\color{black}\textbf{Power}}  & {\color{black}\textbf{Area}} & \textbf{Comment}\\
        \midrule
        This paper   &  MTJ  & $18.3\frac{TOPs}{W}$ & $1.43\frac{TOPs}{W}$  & $3\frac{TOPs}{W}$ & $100$ns/$8.25$ns &  $178$mW / $9.73$mW & $0.068$mm$^2$ & -- \\
       \rule{0pt}{4ex} \multirow{2}{*}{\makecell{\cite{MTJ_BNN_tr} w/o \\ data  converters}} &  \multirow{2}{*}{MTJ} & \multirow{2}{*}{$334\frac{TOPs}{W}$} & \multirow{2}{*}{\makecell{Not\\ evaluated}}  & \multirow{2}{*}{$12.48\frac{TOPs}{W}$} & \multirow{2}{*}{$2$ns/$2$ns} &  \multirow{2}{*}{$22.3$mW/$2.18$mW}& \multirow{2}{*}{\makecell{Not\\ evaluated}} & \multirow{2}{*}{\makecell{Binary weights, FP activations.\\ Original evaluation does\\  not include data converters.}}\\
       \rule{0pt}{9ex} \multirow{2}{*}{\makecell{\cite{MTJ_BNN_tr} w \\data converters}} &  \multirow{2}{*}{MTJ} & \multirow{2}{*}{$0.29\frac{TOPs}{W}$} & \multirow{2}{*}{\makecell{Not\\ evaluated}}  & \multirow{2}{*}{$0.14\frac{TOPs}{W}$} & \multirow{2}{*}{$108$ns/$16.5$ns} &  \multirow{2}{*}{$177$mW/$27.3$mW}& \multirow{2}{*}{\makecell{Not \\evaluated}} & \multirow{2}{*}{\makecell{Evaluation based on\\ Table~\ref{tab:periph}}}\\
        \rule{0pt}{7ex} \cite{Edouard} &  RRAM   & $1326\frac{TOPs}{W}$ & -- & --     & $100$ns/ -- &  $0.25$mW/-- & $0.0012$mm$^2$ & Support Inference of BNN \\
       \rule{0pt}{5ex} \cite{Murmann} & CMOS & $532\frac{TOPs}{W}$  & --   &  --  & -- &  -- & --  &\makecell{Support Inference of BNN \\ Assuming no use of data converters.}\\
        \bottomrule          
    \end{tabular}
    \end{adjustbox}
        \begin{tablenotes}
        \item\small {\color{black}FF, BP, WU are acronyms for feedforward,backpropagation, and weight update, respectively. The delay and power cells format stand for $<$read value$>$/$<$write value$>$.}
      \end{tablenotes}
    \label{tab:Eeff}
\end{table}
\section{Comparison to Previous Work}\label{sec:comp_prev_works}
Most previous work on in-situ hardware implementations of BNN and TNN only support inference. In~\cite{Murmann}, a CMOS-based computation-near-memory engine was designed and fabricated. The design's energy efficiency during inference is $532\frac{TOPs}{W}$. The authors assumed that the binary activation can be done immediately after the convolution, thereby eliminating the ADC. A similar assumption for our setup will increase the inference energy efficiency of our design to $180\frac{TOPs}{W}$. {\color{black}Supporting training in such an accelerator will include additional arithmetic units and will also lead to frequent accesses to the memory to fetch the next layer and will require larger memory capacity to store the intermediate results.}
BNN inference without the need for an ADC is also supported in \cite{Edouard}, where energy efficiency of $1326\frac{TOPs}{W}$ was reported. In that work, an RRAM device was used instead of an MTJ. The RRAM-based synapse can use smaller access transistors than the MTJ-based device, which is current-driven. Moreover, a $1$T$1$R synapse is sufficient when supporting only inference, thereby reducing the complexity and overall power consumption of each synapse.
~\cite{PXNOR-BNN} simulated MTJ-based memory, which supports digital XNOR and XOR operations. By modifying the array drivers, they performed XNOR or XOR operation between operands given to the write driver. The result is written into the memory cell and read by the sense amplifier. Thus, this solution requires three stages to perform the XNOR or XOR operation: preset, XNOR (write), and read, and can perform the operation on a single row each time. They used the MTJ only as a memory cell and did not exploit its stochastic behavior.

Other works exploit the stochastic behavior of the MTJ device.
\cite{MTJ_sto_neuron} exploits the MTJ stochastic switching to design a stochastic neuron. In their work, the training is done off-line, and the weights can be stored in any memristive technology, while the neuron circuit includes an MTJ device.
In~\cite{NV-BNN}, a STT-MTJ-based synapse is used to support BNN training. This solution works in the low current regime; thus, the MTJ switching follows an exponential distribution. Although such distribution is mathematically suitable to train QNNs, our simulations showed that working in the low current regime requires long update periods (approximately $ms$). Our approach is to train in the high current regime, so the stochastic update will occur in a realistic time period.
In \cite{MTJ_BNN_tr}, $1$T$1$R and $1$R structures were proposed, and the stochastic behavior of the MTJ was leveraged to support in-situ training of DNNs with binary weights and full-precision activation. Two update operations are required for the $1$T$1$R and four for the $1$R, whereas our synapse can perform positive and negative operations in parallel, thus requiring only one update operation. 
Moreover, a high-resolution DAC is necessary to convert the full-precision input value to voltages. Using the DAC circuits listed in Table~\ref{tab:periph}, the $8$-bit DAC consumes $5\times$ more energy than the $1$-bit DAC. The power consumption and complexity of their synapse array are, therefore, greater than those of our MTJ-GXNOR. 

Numerous works, which include MTJ-based synapse, focused on the Spike Timing Dependent Plasticity (STDP) learning rule used in bio-inspired ANNs~\cite{Vincent2015,Analitic_MTJ}. Nevertheless, common DNNs are trained with gradient-based optimization~\cite{Ruder}. 
Our work focuses on how the MTJ stochastic behavior can be used to train TNNs and BNNs with a stochastic gradient-based update rule. 

\section{Conclusions}
\label{sec:conclusion}
\par In this paper, we demonstrated the potential of MTJ-based synapse to support in-situ TNN and BNN stochastic training, without sacrificing accuracy. The proposed circuit enables highly parallel and low power execution of weight-related computation. 
We demonstrated its great potential to achieve high energy efficiency in different DNN systems. 
To fulfill the potential of the MTJ-based synapse, the next step is to integrate it into a full system design.

The stochastic behavior of the MTJ can support different training algorithms. For example, while in this work we used MTJ stochastic switching to quantize the gradients, it can be used in algorithms that use stochastic quantization of the weights and activations. Moreover, other optimization algorithms, such as simulated annealing, might benefit from these properties.
The high energy efficiency and the flexibility in functionality enable different algorithms and systems that can accelerate QNN inference and training on low-power devices such as IoT and consumer devices. 

\ack
This work was partially supported by the European Research Council under the European Union’s Horizon 2020 Research and Innovation Program (grant agreement no. 757259) and by the Hasso Plattner Institute for Digital Engineering.

\section*{References}
\bibliography{iopart-num}

\providecommand{\newblock}{}
\begin{thebibliography}{10}
\expandafter\ifx\csname url\endcsname\relax
  \def\url#1{{\tt #1}}\fi
\expandafter\ifx\csname urlprefix\endcsname\relax\def\urlprefix{URL }\fi
\providecommand{\eprint}[2][]{\url{#2}}

\bibitem{Proc_Surv}
{Sze} V, {Chen} Y, {Yang} T and {Emer} J~S 2017 {\em Proceedings of the IEEE\/}
  {\bf 105} 2295--2329

\bibitem{DaDianNao}
Chen Y, Luo T, Liu S, Zhang S, He L, Wang J, Li L, Chen T, Xu Z, Sun N and
  Temam O 2014 {DaDianNao: A Machine-Learning Supercomputer} {\em 2014 47th
  Annual IEEE/ACM International Symposium on Microarchitecture\/} pp 609--622
  ISSN 1072-4451

\bibitem{PipeLayer}
Song L, Qian X, Li H and Chen Y 2017 {PipeLayer: A Pipelined ReRAM-Based
  Accelerator for Deep Learning} {\em 2017 IEEE International Symposium on High
  Performance Computer Architecture (HPCA)\/} pp 541--552

\bibitem{Soudry2014}
Soudry D, Castro D~D, Gal A, Kolodny A and Kvatinsky S 2015 {\em IEEE
  Transactions on Neural Networks and Learning Systems\/} {\bf 26} 2408--2421
  ISSN 2162-237X

\bibitem{Hubara2016}
Hubara I, Courbariaux M, Soudry D, El-Yaniv R and Bengio Y 2016 {\em
  arXiv:1609.07061\/} (\textit{Preprint} \eprint{arXiv:1609.07061v1})

\bibitem{GXNOR}
Deng L, Jiao P, Pei J, Wu Z and Li G 2018 {\em Neural Networks\/} {\bf 100}
  49--58

\bibitem{Courbariaux2016}
Courbariaux M, Hubara I, Soudry D, El-Yaniv R and Bengio Y 2016 {\em
  arXiv:1602.02830\/} ISSN 1664-1078 (\textit{Preprint} \eprint{1602.02830})

\bibitem{Courbariau2015}
Courbariau M, Bengio Y and Davi J~P 2015 {BinaryConnect: Training Deep Neural
  Networks with binary weights during propagations} {\em Advances in Neural
  Information Processing Systems 28\/} pp 3123--3131

\bibitem{XNOR-RRAM}
Sun X, Yin S, Peng X, Liu R, s~Seo J and Yu S 2018 {XNOR-RRAM: A Scalable and
  Parallel Resistive Synaptic Architecture for Binary Neural Networks} {\em
  2018 Design, Automation Test in Europe Conference Exhibition (DATE)\/} pp
  1423--1428

\bibitem{Yu2013}
Yu S, Gao B, Fang Z, Yu H, Kang J and Wong H~S~P 2013 {\em Frontiers in
  Neuroscience\/} {\bf 7} 186 ISSN 1662-453X

\bibitem{Vincent2015}
Vincent A~F, Larroque J, Locatelli N, Romdhane N~B, Bichler O, Gamrat C, Zhao
  W~S, Klein J~O, Galdin-Retailleau S and Querlioz D 2015 {\em IEEE
  Transactions on Biomedical Circuits and Systems\/} {\bf 9} 166--174 ISSN
  19324545

\bibitem{MTJ_BNN_tr}
Mondal A and Srivastava A 2018 {In-situ Stochastic Training of MTJ Crossbar
  Based Neural Networks} {\em Proceedings of the International Symposium on Low
  Power Electronics and Design\/} ISLPED '18 (ACM) pp 51:1--51:6 ISBN
  978-1-4503-5704-3

\bibitem{Ruder}
Ruder S 2016 {\em {CoRR}\/} {\bf abs/1609.04747} (\textit{Preprint}
  \eprint{1609.04747})

\bibitem{MNIST}
Lecun Y, Bottou L, Bengio Y and Haffner P 1998 {\em Proceedings of the IEEE\/}
  {\bf 86} 2278--2324 ISSN 0018-9219

\bibitem{SVHN}
Netzer Y, Wang T, Coates A, Bissacco A, Wu B and Ng A~Y 2011 Reading digits in
  natural images with unsupervised feature learning {\em NIPS Workshop on Deep
  Learning and Unsupervised Feature Learning 2011\/}

\bibitem{CIFAR10}
Krizhevsky A 2012 {\em University of Toronto\/}

\bibitem{ISAAC}
Shafiee A, Nag A, Muralimanohar N, Balasubramonian R, Strachan J~P, Hu M,
  Williams R~S and Srikumar V 2016 {\em 2016 ACM/IEEE 43rd Annual International
  Symposium on Computer Architecture (ISCA)\/}  14--26

\bibitem{Analitic_MTJ}
Vincent A, Locatelli N, Klein J~O, Zhao W, Galdin-Retailleau S and Querlioz D
  2015 {\em IEEE Trans. Electron Devices\/} {\bf 62} 164--170 ISSN 0018-9383

\bibitem{PRNG}
Lambić D and Nikolić M 2017 {Pseudo-random number generator based on
  discrete-space chaotic map} {\em 2014 IEEE International Solid-State Circuits
  Conference Digest of Technical Papers (ISSCC)\/} pp 280--281

\bibitem{Endurance}
Yu S and Chen P~Y 2016 {\em IEEE Solid-State Circuits Magazine\/} {\bf 8}
  43--56

\bibitem{tz_mom}
{Greenberg-Toledo} T, {Mazor} R, {Haj-Ali} A and {Kvatinsky} S 2019 {\em IEEE
  Transactions on Circuits and Systems I: Regular Papers\/} {\bf 66} 1571--1583
  ISSN 1558-0806

\bibitem{Eyal_Sergey}
Rosenthal E, Greshnikov S, Soudry D and Kvatinsky S 2016 {\em Proceedings -
  IEEE International Symposium on Circuits and Systems\/}  1394--1397 ISSN
  02714310

\bibitem{Ms_scale}
Kubota H, Fukushima A, Yakushiji K, Yakata S, Yuasa S, Ando K, Ogane M, Ando Y
  and Miyazaki T 2009 {\em Journal of Applied Physics\/} {\bf 105} 7--117

\bibitem{Gilbert2004}
Gilbert T~L 2004 {\em IEEE Transactions on Magnetics\/} {\bf 40} 3443--3449
  ISSN 0018-9464

\bibitem{GarcuaPalacios1998}
Garc\'{\i}a-Palacios J~L and L\'azaro F~J 1998 {\em Phys. Rev. B\/} {\bf
  58}(22) 14937--14958

\bibitem{Slonczewski1996}
Slonczewski J 1996 {\em Journal of Magnetism and Magnetic Materials\/} {\bf
  159} 1 -- 7 ISSN 0304-8853

\bibitem{Aquino2006}
d’Aquino M, Serpico C, Coppola G, Mayergoyz I~D and Bertotti G 2006 {\em
  Journal of Applied Physics\/} {\bf 99}

\bibitem{Diao2007}
Diao Z, Li Z, Wang S, Ding Y, Panchula A, Chen E, Wang L~C and Huai Y 2007 {\em
  Journal of Physics: Condensed Matter\/} {\bf 19} 165209

\bibitem{Slonczewski1989}
Slonczewski J~C 1989 {\em Phys. Rev. B\/} {\bf 39}(10) 6995--7002

\bibitem{VGG}
Simonyan K and Zisserman A 2014 {\em arXiv preprint arXiv:1409.1556\/}

\bibitem{Li2016}
Li F, Zhang B and Liu B 2016 Ternary weight networks (\textit{Preprint}
  \eprint{1605.04711})

\bibitem{Temp1}
Bi X, Li H and Wang X 2012 {\em IEEE Transactions on Magnetics\/} {\bf 48}
  3821--3824 ISSN 0018-9464

\bibitem{Temp2}
Liu X, Mazumdar D, Shen W, Schrag B~D and Xiao G 2006 {\em Applied Physics
  Letters\/} {\bf 89} 023504

\bibitem{Edouard}
{Giacomin} E, {Greenberg-Toledo} T, {Kvatinsky} S and {Gaillardon} P 2019 {A
  Robust Digital RRAM-Based Convolutional Block for Low-Power Image Processing
  and Learning Applications} vol~66 pp 643--654

\bibitem{DAC}
{Saberi} M, {Lotfi} R, {Mafinezhad} K and {Serdijn} W~A 2011 {\em IEEE
  Transactions on Circuits and Systems I: Regular Papers\/} {\bf 58} 1736--1748
  ISSN 1549-8328

\bibitem{VComp}
{Aiello} O, {Crovetti} P and {Alioto} M 2018 Fully synthesizable, rail-to-rail
  dynamic voltage comparator for operation down to 0.3 v {\em 2018 IEEE
  International Symposium on Circuits and Systems (ISCAS)\/} pp 1--5 ISSN
  2379-447X

\bibitem{Murmann}
{Bankman} D, {Yang} L, {Moons} B, {Verhelst} M and {Murmann} B 2018 {An
  always-on 3.8μJ/86\% CIFAR-10 mixed-signal binary CNN processor with all
  memory on chip in 28nm CMOS} {\em 2018 IEEE International Solid - State
  Circuits Conference - (ISSCC)\/} pp 222--224 ISSN 2376-8606

\bibitem{PXNOR-BNN}
{Chang} L, {Ma} X, {Wang} Z, {Zhang} Y, {Xie} Y and {Zhao} W 2019 {\em {IEEE
  Transactions on Very Large Scale Integration (VLSI) Systems}\/} {\bf 27}
  2668--2679 ISSN 1557-9999

\bibitem{MTJ_sto_neuron}
Zand R, Camsari K~Y, Datta S and Demara R~F 2019 {\em J. Emerg. Technol.
  Comput. Syst.\/} {\bf 15} ISSN 1550-4832

\bibitem{NV-BNN}
{Chang} C, {Wu} M, {Lin} J, {Li} C, {Parmar} V, {Lee} H, {Wei} J, {Sheu} S,
  {Suri} M, {Chang} T and {Hou} T 2019 {NV-BNN: An Accurate Deep Convolutional
  Neural Network Based on Binary STT-MRAM for Adaptive AI Edge} {\em 2019 56th
  ACM/IEEE Design Automation Conference (DAC)\/} pp 1--6 ISSN 0738-100X

\end{thebibliography}

\end{document}